\documentclass[12pt]{article}
\usepackage{url}
\usepackage{amsfonts}
\usepackage{amsmath,amssymb,color}
\usepackage{graphicx}
\usepackage{enumitem}
\usepackage{multirow}
\usepackage{bm}
\usepackage{natbib}
\usepackage{verbatim}
\usepackage{float}
\usepackage[toc,page]{appendix}
\usepackage[onehalfspacing]{setspace}
\usepackage{setspace}
\usepackage{adjustbox}
\usepackage{rotating}
\usepackage{subcaption}
\usepackage{amsthm}
\usepackage{booktabs,dcolumn,caption}

\usepackage{array}
\newcolumntype{L}[1]{>{\raggedright\let\newline\\\arraybackslash\hspace{0pt}}p{#1}}
\newcolumntype{C}[1]{>{\centering\let\newline\\\arraybackslash\hspace{0pt}}p{#1}}
\newcolumntype{R}[1]{>{\raggedleft\let\newline\\\arraybackslash\hspace{0pt}}p{#1}}

\newcommand{\yc}[1]{{{\color{black}  #1}}}

\newcommand{\ttt}{\boldsymbol \theta}

\newcommand{\dd}{\mbox{$\mathbf d$}}

\newcommand{\XX}{\mbox{$\mathbf X$}}

\newcommand{\yy}{\mbox{$\mathbf y$}}

\newcommand{\aaaa}{\mathbf a}

\newcommand{\xx}{\mathbf x}

\newcommand{\argmax}{\operatornamewithlimits{arg\,max}}
\newcommand{\argmin}{\operatornamewithlimits{arg\,min}}

\newtheorem{theorem}{Theorem}

\newtheorem{algorithm}{Algorithm}

\newtheorem{corollary}{Corollary}

\title{Joint Maximum Likelihood Estimation for High-dimensional Exploratory Item Factor Analysis}
\author{Yunxiao Chen\\
 Department of Statistics, London School of Economics and Political Science\\
 Xiaoou Li\\
 School of Statistics, University of Minnesota\\
 Siliang Zhang\\
 Shanghai Center for Mathematical Sciences, Fudan University
 }
\date{}

\begin{document}
\maketitle

\doublespacing

\begin{abstract}
\yc{Joint maximum likelihood (JML) estimation is one of the earliest approaches to fitting item response theory (IRT) models.
This procedure treats both the item and person parameters as unknown but fixed model parameters, and estimates them simultaneously by solving an optimization problem. However, the JML estimator is known to be asymptotically inconsistent for many IRT models, when the sample size goes to infinity and the number of items keeps fixed.  Consequently, in the psychometrics literature, this estimator is less preferred to the marginal maximum likelihood (MML) estimator. In this paper, we re-investigate the JML estimator for high-dimensional exploratory item factor analysis, from both statistical and computational perspectives. In particular, we
establish a notion of statistical consistency for a constrained JML estimator,
under an asymptotic setting that both the numbers of items and people grow to infinity and that many responses may be missing. A parallel computing algorithm is proposed for this estimator that can scale to very large datasets.
Via simulation studies, we show that when the dimensionality is high, the proposed estimator yields similar or even better results than those from the MML estimator, but can be obtained computationally much more efficiently. An illustrative real data example  is provided based on the revised version of Eysenck's Personality Questionnaire (EPQ-R).}


\end{abstract}	

\clearpage

\section{Introduction}\label{Sec:intro}

Exploratory Item Factor Analysis \citep[IFA;][]{bock1988full} has been widely used as an analytic approach to analyzing item-level data within social and behavioral sciences \citep{bartholomew2008analysis}. Such data are typically either dichotomous (e.g., disagree vs. agree) or polytomous (e.g., strongly disagree, disagree, neither, agree, and strongly agree), for which the standard linear factor models  may not be suitable \citep{wirth2007item}.
Exploratory IFA uncovers and interprets the underlying structure of data by learning the association between the
items and the latent factors based on the estimated factor loadings. It has received many applications in social and behavioral sciences, including but not limited to personality, quality-of-life, and clinical research \citep[e.g.,][]{edelen2007applying,Lee2009factor,reise2009item}.

There are a wide range of psychometric models for exploratory item factor analysis.
For the purpose of exploratory analysis, all these models handle multiple latent factors,
including Multidimensional Two-parameter Logistic Model \citep[M2PL;][]{Reckase1972,reckase2009multidimensional} for dichotomous responses, the multidimensional graded response model \citep[e.g.,][]{cai2010high} and multidimensional partial credit model \citep{yao2006multidimensional} for polytomous responses, and normal ogive (i.e., probit) models for
dichotomous and polytomous responses \citep{bock1988full}. The readers are referred to \cite{wirth2007item} for a comprehensive review of the IFA literature. For ease of exposition, we focus on IFA models for dichotomous responses, while point out that our developments can be extended to polytomous data.


The most commonly used method for parameter estimation in exploratory IFA is marginal maximum likelihood (MML) estimation based on an Expectation-Maximization (EM) algorithm \citep{bock1981marginal,bock1988full}.
In this approach, the item parameters are estimated by maximizing the marginal likelihood function, in which the person parameters (i.e., latent factors) have been integrated out. This approach typically involves evaluating a $K$-dimensional integral, where $K$ is the number of latent factors.
The computational complexity of evaluating this integral grows exponentially with the latent dimension $K$ and the computation becomes infeasible when the latent dimension is too high.
In fact, the Gauss-Hermite quadrature-based integration used by \cite{bock1981marginal} is not recommended for more than five factors \citep{wirth2007item}, which limits the use of MML estimation in large-scale data analysis where many latent factors may be present.
In filling this gap, many approaches have been proposed to approximate the integral,
including adaptive Gaussian quadrature methods \citep[e.g.][]{schilling2005high},
Monte Carlo integration \citep[e.g.][]{meng1996fitting}, fully Bayesian estimation methods \citep[e.g.][]{beguin2001mcmc,bolt2003estimation,edwards2010markov},
and data augmented stochastic approximation algorithms \citep[e.g.][]{cai2010high, cai2010metropolis}.
However, even with these state-of-the-art algorithms, the computation is time-consuming with the presence of many latent factors.

Alternative approaches have been proposed for parameter estimation in IFA that avoid evaluating high-dimensional integrals. These approaches are computationally more efficient and thus may be more suitable for the analysis of large-scale data. 
In particular, \cite{lee1990three} propose to first estimate the inter-item polychoric correlation matrix using pairwise response data and then to estimate the loadings by conducting factor analysis based on the estimated polychoric correlation matrix. However, this approach relies heavily on the assumptions of normal ogive models and can hardly be generalized when other link functions are used. \cite{joreskog2001factor} propose a composite likelihood approach that maximizes the sum of all
univariate and bivariate marginal likelihoods. In this approach, only one- and two-dimensional numerical integrals need to be evaluated, which is computationally more affordable than that of the MML approach. However, this approach still relies heavily on the assumption that the latent factors follow a multivariate normal distribution which may not always be satisfied in applications.

Joint maximum likelihood (JML) estimator is one of the earliest approaches to parameter estimation for IFA models that is known to be computationally efficient \citep[see Chapter 8,][]{embretson2000item}. This approach is first suggested in \cite{birnbaum1968some} when the basic forms of item response theory models are proposed and has been used in item response analysis for many years \citep{lord1980applications,mislevy1987consumer} until the MML approach becomes dominant. The key difference between the MML and the JML methods is that, in the MML approach the person parameters are treated as random effects and are integrated out from the likelihood function, while in the JML approach the person parameters are treated as fixed effect parameters and kept in the likelihood function. As a result, the evaluation of numerical integrals in the MML approach is replaced by maximizing with respect to the person parameters in the JML approach.
Under a latent factor model with a high latent dimension,
the computational complexity of the latter is much lower than that of the former.
However, in the IFA literature,  JML estimation is less preferred to MML estimation.
This is because, under the classical asymptotic setting where the number of respondents grows to infinity and the number of items is fixed, the number of parameters in the joint likelihood function also grows to infinity, for which the standard theory for maximum likelihood estimation does not apply. Consequently, the point estimation of every single item parameter is inconsistent \citep{neyman1948consistent,andersen1973conditional,haberman1977maximum,ghosh1995inconsistent} even for simple IRT models, let alone the validity of the standard errors for the item parameter estimates.

\yc{Despite its statistical inconsistency in the classical sense, the JML approach is computationally efficient, easily programmable, and generally applicable to many IRT models \citep{embretson2000item}. Though possibly biased, the empirical performance of JML estimator for point estimation is usually reasonable, especially when constraints are placed on the JML solution. Given the unique strength of JML-based estimation,  its properties are worth investigating from a theoretical perspective.}
In this paper, we provide statistical theory to IFA based on the joint likelihood for analyzing large-scale data where both the number of people and the number of items are large. Our asymptotic setting differs from the standard one by letting both the numbers of people and items grow to infinity. This setting seems reasonable for analyzing large-scale item response data. Similar asymptotic settings have been considered in psychometric research, including the analysis of unidimensional IRT models \citep{haberman1977maximum,haberman2004joint}
and  diagnostic classification models \citep{chiu2016joint}.
Under this asymptotic setting, we propose a constrained joint maximum likelihood estimator (CJMLE) that has certain notion of statistical consistency in recovering factor loadings. \yc{Since the number of loading parameters grows to infinity under this asymptotic setting, this notion of consistency is different from that in the classical sense for maximum likelihood estimation. Specifically, we show that, up to a rotation, the proportion of inconsistently estimated loading parameters converges to zero in probability. }



The major advantage of the proposed CJMLE over the MML-based approaches is its low computational cost. An alternating minimization (AM) algorithm with projected gradient decent update is proposed, which can be parallelled efficiently. Specifically, we implement this parallel computing algorithm in R with core functions written in C++ through Open Multi-Processing \citep[OpenMP,][]{dagum1998openmp} that can scale to very large data. For example,  the algorithm can fit a dataset with 125,000 respondents, 500 items, and 10 latent traits within 3 minutes on a single Intel$^\circledR$ machine\footnote{Core(TM) i7CPU@5650U@2.2 GHz.} with four cores.
Compared with \cite{lee1990three} and \cite{joreskog2001factor},
our method is not only more flexible for its ability to handle almost all IFA models, but also computationally more efficient. Specifically,   the computational complexity of our method is linear in the number of items while that of \cite{lee1990three} and \cite{joreskog2001factor} is quadratic.


\yc{As an illustration, we apply the proposed estimator to a personality assessment dataset based on a revised version of the Eysenck's personality questionnaire \citep{eysenck1985revised}. This dataset contains 79 items, which are designed to measure three personality factors, Extraversion (E), Neuroticism (N), and  Psychoticism (P). It is found that a three-factor model fits the data best, according to a cross-validation procedure. In addition, the three factors identified by the Geomin rotation \citep{yates1988multivariate} correspond well to the three factors in Eysenck's model of personality.}

The remainder of the paper is organized as follows. In Section~\ref{Sec:method}, we propose the constrained joint maximum likelihood estimator under a general form of IFA models and establish its asymptotic properties. Then in Section~\ref{Sec:compute}, a computational algorithm is proposed. Simulation studies and real data analysis are presented in in Sections \ref{Sec:simulation} and \ref{Sec:real}, respectively. Finally, discussions are provided in Section~\ref{Sec:disc}. Proofs of our theoretical results are provided in supplementary material.

\section{Constrained Joint Maximum Likelihood Estimation}\label{Sec:method}

\subsection{IFA Models for Dichotomous Responses}

We focus on a class of IFA models for dichotomous responses, which includes the M2PL model and the normal ogive model as special cases. Let $i = 1, ..., N$ indicate respondents and $j = 1, ..., J$
indicate items. Each respondent $i$ is represented by a $K$-dimensional latent vector $\ttt_i = (\theta_{i1}, ..., \theta_{iK})^\top$ and each item is represented by $K+1$ parameters including an intercept parameter $d_j$ and $K$ loading parameters $\aaaa_j = (a_{j1}, ..., a_{jK})^\top$. Let $Y_{ij}$ be the response from respondent $i$ to item $j$, which is assumed to follow distribution
\begin{equation}\label{eq:model}
P(Y_{ij} = 1 \vert \ttt_i, d_j, \aaaa_j) = f(d_j + \aaaa_j^\top \ttt_i),
\end{equation}
where $f(x)$ is a pre-specified link function. \yc{Given the latent vector $\ttt_i$, respondent $i$'s responses $Y_{i1}, ..., Y_{iJ}$ are assumed to be conditionally independent. This assumption is known as the local independence assumption, a standard assumption for item factor analysis.} We denote the observed value of $Y_{ij}$ by $y_{ij}$.

The framework \eqref{eq:model} includes the M2PL model and the normal ogive model as special cases. Specifically, for the M2PL model, the link function takes the logistic form
$$f(x) = \frac{\exp(x)}{1+\exp(x)},$$
and for the normal ogive model, the link function becomes
$$f(x) = \int_{-\infty}^x \phi(t)dt,$$
where $\phi(x)$ is the probability density function of a standard normal distribution.
Besides these two widely used models, other link functions may also be used, such as a complimentary
log-log link or a link function with pre-specified lower and/or upper asymptotes. 

Given a model, the MML-based IFA further requires the specification of a prior distribution on the latent factors $\ttt_i$. \yc{In fact, the consistency of MML estimation relies on the correct specification of the prior distribution, under the classical asymptotic setting.}
For exploratory IFA, a commonly adopted assumption is that $\ttt_i$ follows a $K$-dimensional standard normal distribution. In the implementation of the Gauss-Hermite quadrature-based EM algorithm, this distribution is further approximated by a discrete distribution supported on a ball.
In contrast, as will be described in the sequel, the JML-based IFA does not require the specification of a prior.

\subsection{Constrained Joint Maximum Likelihood Estimation}

Under the general model form~\eqref{eq:model}, the joint likelihood function is a function of both the item parameters $\aaaa_j$ and $d_j$ and the person parameters $\ttt_i$, specified as

\begin{equation}\label{eq:Jlik}
\begin{aligned}
&L(\ttt_i, \aaaa_j, d_j: i = 1, ...,N, j = 1, ..., J ) \\
=& \prod_{i=1}^N \prod_{j=1}^J f(d_j + \aaaa_j^\top \ttt_i)^{y_{ij}} (1-f(d_j + \aaaa_j^\top \ttt_i))^{1-y_{ij}}.
\end{aligned}
\end{equation}
The classical JML estimator is defined as the maximizer of the joint likelihood function
\begin{equation}\label{eq:jml}
\begin{aligned}
&(\hat \ttt_i, \hat \aaaa_j, \hat d_j: i = 1, ...,N, j = 1, ..., J) \\
=&  \argmax_{\ttt_i, \aaaa_j, d_j} \log L(\ttt_i, \aaaa_j, d_j: i = 1, ...,N, j = 1, ..., J).
\end{aligned}
\end{equation}

One issue with the JML estimator is that estimates are not available for items or persons with perfect scores (all 1s or all 0s), when no constraints are placed. To avoid this issue, we propose a constrained joint maximum likelihood estimator (CJMLE), defined as
\begin{equation}\label{eq:rjmle}
\begin{aligned}
&(\hat \ttt_i, \hat \aaaa_j, \hat d_j: i = 1, ...,N, j = 1, ..., J) \\
=&  \argmax_{\ttt_i, \aaaa_j, d_j} \log L(\ttt_i, \aaaa_j, d_j: i = 1, ...,N, j = 1, ..., J)\\
s.t. ~& \sqrt{1+\Vert \ttt_{i} \Vert^2} \leq C,  \sqrt{d_j^2 + \Vert \aaaa_{j} \Vert^2} \leq C,  i = 1, ...,N, j = 1, ..., J.
\end{aligned}
\end{equation}
Throughout this paper, $\Vert\xx\Vert$ denotes the Euclidian norm of a vector $\xx = (x_1, ..., x_K)$, defined as $\Vert\xx\Vert = \sqrt{x_1^2 + x_2^2 + \cdots x_K^2}$. In \eqref{eq:rjmle}, $C$ is a pre-specified positive constant that imposes regularization on the magnitudes of the person-wise parameters and the item-wise parameters. \yc{Since the feasible set given by the constraints in \eqref{eq:rjmle} is compact and the objective function is continuous, the optimization problem is guaranteed to have a solution. Therefore, estimates exist even for items and persons with perfect scores. It is also worth pointing out that the solution to \eqref{eq:rjmle} is not unique, due to rotational indeterminacy \citep{browne2001overview}, to be further discussed in Section~\ref{subsec:theory_loading}.}
As will also be shown in Section~\ref{subsec:theory_loading}, the CJMLE has statistical guarantees for any sufficiently large value of $C$, under the asymptotic regime where both $N$ and $J$ grow to infinity. In the rest of the paper, we use $C = 5\sqrt{K}$ as a default value under the M2PL model. 

\subsection{Theoretical Properties: Recovery of Response Probabilities}\label{subsec:theory_probability}

We establish the asymptotic properties of the CJMLE defined in \eqref{eq:rjmle}. We denote $\ttt_i^*$, $\aaaa_j^*$, and $d_j^*$ the true model parameters, where $i = 1, 2, ..., N$, $j = 1, 2, ..., J$. In this analysis, the dimension $K$ of the latent space is known, while in practice one may choose a dimension $K$ either via cross-validation or by using an information criterion. We introduce the following notations.
\begin{enumerate}
  \item $\Theta = (\theta_{ik})_{N\times K}$ denotes the matrix of person parameters.
  \item $A = (a_{jk})_{J\times K}$ denotes the matrix of factor loadings.
  \item $\mathbf d = (d_1, ..., d_J)$ denotes the vector of  intercept parameters.
  \item $\Theta^* = (\theta_{ik}^*)_{N\times K}$, $A^* = (a_{jk}^*)_{J\times K}$ and $\mathbf d^*$ denote the true parameters.
  \item $\mathbf 1_N = (1, ..., 1)$ denotes a vector with all $N$ entries being 1.
  \item $X_{[k]}$ denotes the $k$th column vector of a matrix $X$.
  \item $\hat \Theta = (\hat \theta_{ik})_{N\times K}$, $\hat{\mathbf d} = (\hat  d_1, ..., \hat d_J)$, and
  $\hat A = (\hat a_{jk})_{J \times K}$ denote the CJMLE given in \eqref{eq:rjmle}.
  \item $\Vert X\Vert_F = \sqrt{\sum_{i=1}^N\sum_{j=1}^J x_{ij}^2}$ denotes the Frobenius norm of a matrix $X = (x_{ij})_{N\times J}$.
\end{enumerate}
In addition, we require the following regularity conditions.
\begin{enumerate}
  \item[A1.] $\sqrt{1+\Vert \ttt_{i}^* \Vert^2} \leq C$ and  $\sqrt{(d_j^*)^2 + \Vert \aaaa_{j}^* \Vert^2} \leq C,  i = 1, ...,N, j = 1, ..., J.$
  \item[A2.] The link function $f$ is differentiable, satisfying
  $$\sup_{\vert x\vert \leq C^2} \frac{\vert f'(x)\vert}{f(x)(1-f(x))} <  \infty \mbox{~~and~~} \sup_{\vert x\vert \leq C^2} \frac{f(x)(1-f(x))}{(f'(x))^2} < \infty.$$
\end{enumerate}
These two conditions are reasonable and easy to understand. Condition A1 requires that the true person parameters and the true item parameters satisfy the constraints used in the CJMLE defined in \eqref{eq:rjmle}. Condition A2 requires that the link function $f$ has a certain level of smoothness. In particular, the commonly used link functions, including the logit, probit, and the complimentary log-log links, satisfy A2.

\begin{theorem}\label{thm:concen}
Suppose that assumptions A1 and A2 are satisfied. Then there exist constants $C_1$ and $C_2$ that depend on the value of $C$ (but independent of $N$ and $J$), such that
\begin{equation}\label{eq:thm}
\frac{1}{NJ} \Vert \hat \Theta \hat A^\top + \mathbf 1_N \hat{\mathbf d}^\top  - \Theta^*  (A^*)^\top  - \mathbf 1_N{\mathbf d^*}^\top \Vert^2_F \leq C_2 \sqrt{\frac{J+N}{NJ}}
\end{equation}
is satisfied with probability at least $1 - C_1/(N+J)$, \yc{where $\Vert \cdot\Vert_F$ denotes the matrix Frobenius norm defined above}.
\end{theorem}
The proof of Theorem~\ref{thm:concen} is given in the supplementary material that makes use of a concentration inequality proved in \cite{davenport20141}. The bound \eqref{eq:thm} is satisfied for all $N$ and $J$, without requiring $N$ and $J$ to grow to infinity. When both $N$ and $J$ grow to infinity, Theorem~\ref{thm:concen} implies that the left side of \eqref{eq:thm} converges to 0 in probability.

Theorem~\ref{thm:concen} is essentially about the accuracy of estimating the true response probabilities.
This is because the conditional distribution of $Y_{ij}$ depends on $\ttt_i$, $\aaaa_j$, and $d_j$ only through $d_j + \aaaa_j^\top \ttt_i$, the $(i,j)$th entry of the matrix $\Theta A^\top + \mathbf 1_N {\mathbf d}^\top$. Consequently, the left side of \eqref{eq:thm} quantifies an averaged discrepancy between the true values $\Theta^*  (A^*)^\top  + \mathbf 1_N{\mathbf d^*}$ and their estimates $\hat \Theta \hat A^\top + \mathbf 1_N \hat{\mathbf d}^\top$.
Moreover, Theorem~\ref{thm:concen} implies the consistent recovery of the response probabilities in an average sense, as described in Corollary~\ref{cor:recoveryP}.
\begin{corollary}[Recovery of Response Probabilities]\label{cor:recoveryP} Under the same conditions as Theorem~\ref{thm:concen}, when $N$ and $J$ grow to infinity,
\begin{equation}\label{eq:recoveryP}
\frac{\sum_{i=1}^N\sum_{j=1}^J \left(f(\hat d_j + \hat \aaaa_j^\top \hat \ttt_i) - f(d_j^* + (\aaaa_j^*)^\top \ttt_i^*)\right)^2}{NJ}
\end{equation}
converges to zero in probability.
\end{corollary}
Note that $f(\hat d_j + \hat \aaaa_j^\top \hat \ttt_i)$ is the predicted probability of $Y_{ij} = 1$ given by the CJMLE and $f(d_j^* + (\aaaa_j^*)^\top \ttt_i^*)$ is the corresponding true probability. Therefore, the result of Corollary~\ref{cor:recoveryP} implies that the predicted probabilities and their true values are close in an average sense.
\yc{It further implies that only a small proportion of true item response probabilities are not estimated well; that is, for any small constant $\epsilon >0$, the proportion
$$\frac{\sum_{i=1}^N \sum_{j=1}^J 1_{\{\vert f(\hat d_j + \hat \aaaa_j^\top \hat \ttt_i) - f(d_j^* + (\aaaa_j^*)^\top \ttt_i^*) \vert > \epsilon \}}}{NJ}$$
converges to zero in probability.
This property may be important to psychological measurement, as the item response probabilities completely characterize the respondents' behavior on the items. 

To our knowledge, the type of asymptotic result established in Corollary~\ref{cor:recoveryP} is not considered in the classical asymptotic theory based on the marginal maximum likelihood. In fact, under the classical asymptotic setting,  the quantity \eqref{eq:recoveryP} does not converge to zero in probability if the number of items $J$ is fixed, no matter how the parameters are estimated.
}

\subsection{Theoretical Properties: Recovery of Loadings}\label{subsec:theory_loading}

We now study the recovery of the loading structure $A^*$, which is of particular interest in exploratory IFA.
Specifically, we will show that $\hat A$ given by the CJMLE approximates $A^*$ well in a sense to be clarified.

We start the discussion with the identifiability of the model parameters. Given all the true response probabilities, or equivalently, the matrix $\Theta^*  (A^*)^\top  + \mathbf 1_N{\mathbf d^*}^\top$, the parameters $\Theta^*$, $A^*$, and $\mathbf d^*$ cannot be uniquely determined. To avoid this indeterminacy issue, we impose the following regularity condition on the true person parameters.
\begin{enumerate}
  \item[A3.] The true person parameters satisfy
\begin{align}
& \mathbf 1_N^\top \Theta_{[k]}^* = 0, \label{eq:identify1}\\
& \frac{1}{N}(\Theta_{[k]}^*)^\top\Theta_{[k]}^* = 1, \label{eq:identify2}\\
& (\Theta_{[k]}^*)^\top\Theta_{[k']}^* = 0, ~~  k, k' = 1, ..., K, k\neq k'. \label{eq:identify3}
\end{align}
\end{enumerate}

The constraints in A3 are similar to assuming the means and covariance matrix of $\ttt_i$ are 0s and identity matrix, respectively, when analyzing data using an MML approach. Even under these constraints, $\Theta^*$ and $A^*$ are only determined up to a rotation, known as rotational indeterminacy. A summary of the phenomenon of rotational indeterminacy is given in the supplementary material.




Taking the constraints \eqref{eq:identify1}-\eqref{eq:identify3} into account, we standardize the CJMLE solution $(\hat \Theta, \hat A, \hat \dd)$, so that the same constraints are satisfied. The standardized solution is denoted by $(\tilde \Theta, \tilde A, \tilde \dd)$, where the standardization procedure is given in the supplementary material.
We then show that $\tilde A$ accurately estimates $A^*$ up to a rotation, when the following regularity condition also holds.
\begin{enumerate}
  \item[A4.] There exists positive constants $C_3 > 0$, such that the $K$th (i.e., the smallest) singular value of $A^*$, denoted by $\sigma_K^*$, satisfying
      $\sigma_{K}^* \geq C_3 \sqrt{J},$
     for all $J$.
\end{enumerate}

\begin{theorem}\label{thm:perturb}
Suppose that assumptions A1 - A4 are satisfied.  Then the following scaled Frobenius loss
\begin{equation}\label{eq:loss}
\min_Q\left\{\frac{1}{JK}\Vert A^* - \tilde A Q\Vert_F^2: Q^\top Q = I_{K\times K} \right\}
\end{equation}
converges to zero in probability as $N,J\to\infty$,
where
$\tilde A$ is the standardized version of $\hat A$.
\end{theorem}

%



\yc{We remark on the result of Theorem~\ref{thm:perturb}. Suppose that $\tilde Q$ minimizes the optimization problem~\eqref{eq:loss}. In addition, we denote $\bar{A} = (\bar a_{jk})_{J\times K} = \tilde A \tilde Q$. Then \eqref{eq:loss} converging to zero implies that for any $\epsilon > 0$,
$$\lim_{N, J \rightarrow \infty} \frac{\sum_{j=1}^J \sum_{k=1}^K 1_{\{\vert a_{jk}^* -  \bar a_{jk} \vert > \epsilon\}}}{JK} = 0.$$
That is, the proportion of inaccurately estimated loading parameters converges to zero in probability under the optimal rotation. }

\yc{In practice, the optimal rotation $\tilde Q$ is not available, since $A^*$ is unknown. A suitable rotation may be obtained by using analytic rotation methods \citep[see e.g.][]{browne2001overview} to yield  a simple pattern of factor loadings that is easy to interpret, where a simple loading pattern refers to a loading matrix with many entries close to 0, so that each item is mainly associated with a small number of latent factors and each latent factor is mainly associated with a small number of items. When the true loading matrix $A^*$ has a simple pattern, we believe that a certain notion of consistency can be established for analytic rotation methods. }

Finally, we remark that condition A4 is mild. In fact, when $\aaaa_j^*$s are i.i.d. random vectors from a distribution and the covariance matrix of $\aaaa_j^*$ is strictly positive definite, $\sigma_{K}^* \geq C_3 \sqrt{J}$ is satisfied with probability close to 1 for sufficiently large $J$, when taking $C_3$ to be $0.5\sqrt{\lambda_K}$, where $\lambda_K$ is the smallest eigenvalue of the covariance matrix of $\aaaa_j^*$.

%
%
%
%

\subsection{Extension: Analyzing Missing Data}\label{subsec:missing}

In practice, each respondent may only respond to a small proportion of items, possibly due to the data collection design.
The proposed CJMLE also handles missing data. More precisely, let $W_{ij}$ indicate whether or not the $(i,j)$th entry of the response matrix is missing, where $W_{ij} = 0$ if the corresponding  response is missing and $W_{ij} = 1$  otherwise. We say the missingness is ignorable when the following equation holds
\begin{equation*}
  \begin{aligned}
     & P(Y_{i1} = y_1, ..., Y_{iJ}=y_J, W_{i1} = \omega_1, ..., W_{iJ} = \omega_J \vert \ttt_i, \aaaa_j, d_j) \\
     =& P(Y_{i1} = y_1, ..., Y_{iJ}=y_J \vert \ttt_i, \aaaa_j, d_j) \times P(W_{i1} = \omega_1, ..., W_{iJ} = \omega_J \vert \ttt_i, \aaaa_j, d_j)\\
     =& \left(\prod_{j=1}^J P(Y_{ij} = y_j\vert \ttt_i, \aaaa_j, d_j)\right)\times \left(\prod_{j=1}^J P(W_{ij} = \omega_j\vert \ttt_i, \aaaa_j, d_j)\right).
  \end{aligned}
\end{equation*}
Let $\omega_{ij}$ be a realization of $W_{ij}$. Then the responses $y_{ij}$ are only observed for the entries with
$\omega_{ij} = 1$. Under ignorable missingness, the joint likelihood function becomes
\begin{equation}\label{eq:Jlik2}
\begin{aligned}
&L(\ttt_i, \aaaa_j, d_j: i = 1, ...,N, j = 1, ..., J) \\
=& \prod_{i,j: \omega_{ij} = 1} f(d_j + \aaaa_j^\top \ttt_i)^{y_{ij}} (1-f(d_j + \aaaa_j^\top \ttt_i))^{1-y_{ij}}.
\end{aligned}
\end{equation}
When $\omega_{ij} = 1$ for all $i$ and $j$, no response is missing and \eqref{eq:Jlik2} becomes the same as \eqref{eq:Jlik}.

The statistical guarantee established earlier for complete data can be extended to data with massive missingness. For technical simplicity, we assume that the data are missing completely at random.
\begin{enumerate}
  \item[A5.] $W_{ij}$s  are i.i.d. Bernoulli random variables with
  $$P(W_{ij} = 1) = \frac{n}{NJ},$$
  for some $n > 0$.
\end{enumerate}
Under this assumption, Theorems~\ref{thm:concen2} and \ref{thm:perturb2} extend Theorems~\ref{thm:concen} and \ref{thm:perturb}  by allowing for missing data. In fact, Theorems~\ref{thm:concen} and \ref{thm:perturb} can be viewed as special cases of Theorems~\ref{thm:concen2} and \ref{thm:perturb2} when $n = NJ$.  The proofs of Theorems~\ref{thm:concen2} and \ref{thm:perturb2} are given in the supplementary material.

\begin{theorem}\label{thm:concen2}
Suppose that assumptions A1, A2, and A5 are satisfied. Further assume that $n \geq (N + J)\log(JN).$ Then there exist constants $C_1$ and $C_2$ that depend on the value of $C$ (but independent of $N$ and $J$), such that
\begin{equation}\label{eq:thm2}
\frac{1}{NJ} \Vert \hat \Theta \hat A^\top + \mathbf 1_N \hat{\mathbf d}^\top  - \Theta^*  (A^*)^\top  - \mathbf 1_N{\mathbf d^*}^\top \Vert^2_F \leq C_2 \sqrt{\frac{J+N}{n}}
\end{equation}
is satisfied with probability at least $1 - C_1/(N+J)$.
\end{theorem}

\begin{theorem}\label{thm:perturb2}
Suppose that assumptions A1 - A5 are satisfied.  Further assume that $n \geq (N + J)\log(JN).$ Then the following scaled Frobenius loss
\begin{equation}\label{eq:loss2}
\min_Q\left\{\frac{1}{JK}\Vert A^* - \tilde A Q\Vert_F^2: Q^\top Q = I_{K\times K} \right\}
\end{equation}
converges to zero in probability as $N,J\to\infty$, where
$\tilde A$ is the standardized version of $\hat A$.
\end{theorem}


Noting that when $n \geq (N + J)\log(JN)$, the right side of equations~\eqref{eq:thm2}
converges to zero when $N$ and $J$ grow to infinity. Consequently, Corollary~\ref{cor:recoveryP} can be extended to this missing data setting. \yc{This asymptotic validity of the CJMLE for missing data suggests its potential in applications of test equating and linking, which can be formulated into missing data analysis problems \citep[see e.g.,][]{von2010statistical}.

We provide a discussion on condition A5. Under certain data collection designs, data can be regarded as missing completely at random (MCAR).  However, it is often the case in practice that the MCAR assumption may be too strong. Instead, it may be more reasonable to assume missing at random (MAR), under which the probability of observing a response $P(W_{ij} = 1)$ depends on the corresponding parameter values, including $\ttt_i$, $\aaaa_j$, and $d_j$. Our theoretical results may be extended to the MAR setting (e.g., using techniques from \citealp{cai2013max}).
}

\subsection{Selection of Number of Factors}\label{subsec:numFac}

We provide a cross-validation method for the selection of the number $K$ of latent factors when it is unknown.
Let $\Omega = (\omega_{ij})_{N\times J}$ be the indicator matrix of non-missing responses. We randomly split the non-missing responses into $B$ non-overlapping sets that are of equal sizes, indicated by $\Omega^{(b)} = (\omega^{(b)}_{ij})_{N\times J}, b = 1, 2, ..., B$,
satisfying $\sum_{b=1}^B \Omega^{(b)} = \Omega$.
Moreover, we denote $\Omega^{(-b)} = \sum_{b'\neq b} \Omega^{(b')}$, indicating the data excluding set $b$.

For a given latent dimension $K$, we find the CJMLE based on the non-missing responses indicated by $\Omega^{(-b)}$. The CJMLE solution is denoted by $(\hat \Theta^{(b)}, \hat A^{(b)}, \hat{\mathbf d}^{(b)})$.  As defined below, the cross-validation error for fold $b$ is computed based on the accuracy of predicting the responses in the set $\Omega^{(b)}$ using $(\hat \Theta^{(b)}, \hat A^{(b)}, \hat{\mathbf d}^{(b)})$.
$$err^{(b)}(K) = \sum_{i,j: ~\omega_{ij}^{(b)} = 1} \left(y_{ij} -  f(\hat{d}_j^{(b)} + (\hat\aaaa_j^{(b)})^\top \hat \ttt_i^{(b)})\right)^2.$$
The overall cross validation error is defined as
$$err(K) = \sum_{b=1}^B err^{(b)}(K).$$
The latent dimension $K$ that yields the smallest cross validation error is selected. In the analysis of this paper, we set $B=5$ (i.e., five-fold cross validation).

\section{Computation}\label{Sec:compute}


We develop an alternating minimization algorithm for solving the optimization problem~\eqref{eq:rjmle}.
\yc{In fact, the first JML estimation paradigm employed in \cite{birnbaum1968some}  can be regarded as an alternating minimization algorithm. This paradigm is the basis for JML estimation for many IRT computer programs in general use \citep{baker1987methodology}.
As indicated by its name, this algorithm decomposes the parameters into two sets, the person parameters and the item parameters, and alternates between minimizing one set of parameters given the other.
It is worth noting that given the person parameters, the optimization with respective to item parameters can be split into $J$ independent optimization problems, each containing $(\aaaa_j, d_j)$, $j = 1, ..., J$. Similarly, the person parameters can also be updated independently for $\ttt_i$, $i = 1, ..., N$, given the item parameters. }

To handle the constraints in
\eqref{eq:rjmle}, a projected gradient descent update is used in each iteration, defined as follows.
We first define projection operator
\begin{equation}\label{eq:proj}
\text{Prox}_{C}(\yy) = \argmin_{\xx: \Vert \xx\Vert \leq C}   \Vert \yy - \xx \Vert^2 =\left\{ \begin{array}{ll}
                                                                                                      \yy &\text{if } \Vert\yy\Vert \leq C;  \\
                                                                                                      \frac{C}{\Vert \yy\Vert}\yy&\text{if } \Vert\yy\Vert > C.\\
                                                                                                      \end{array}\right.
\end{equation}
Here, $\text{Prox}_{C}(\yy)$ returns the projection of $\yy$ onto the feasible set.
Consider optimization problem
\begin{equation}\label{eq:proximal}
\begin{aligned}
 \min_{\xx}~& f(\xx)\\
 s.t.~ & \Vert \xx\Vert \leq C,
\end{aligned}
\end{equation}
where $f$ is a differentiable convex function. Denote the gradient of $f$ by $g$.  Then a projected gradient descent update at $\xx^{(0)}$ is
defined as
$$\xx^{(1)} = \text{Prox}_{C}(\xx^{(0)} - \eta g(\xx^{(0)})),
$$
where $\eta > 0$ is a step size decided by line search. Due to the projection, $\Vert \xx^{(1)} \Vert\leq C$.
Furthermore, it can be shown that for sufficiently small $\eta$,
$f(\xx^{(1)}) <  f(\xx^{(0)})$, when $f$ satisfies mild regularity conditions and $\Vert g(\xx^{(0)}) \Vert \neq 0$;
see \cite{parikh2014proximal} for further details.

\begin{algorithm}[Alternating Minimization Algorithm for CJMLE]\label{alg:am}

~

\begin{enumerate}
\item[1] (Initialization) Input responses $y_{ij}$, nonmissing response indicator $\omega_{ij}$, dimension $K$ of latent space, constraint parameter $C$,
iteration number $m = 0$, and initial value $\Theta^{(0)}$, $A^{(0)}$, $\dd^{(0)}$.
\item[2] (Alternating minimization) At the $m+1$th iteration,

\begin{itemize}
  \item[(a)] Perform parallel computation for $i = 1, ..., N$.
  For each respondent $i$, update
    $\ttt_i^{(m+1)} = \text{Prox}_{\sqrt{C^2-1}}\left(\ttt_i^{(m)} - \eta \mathbf g_i^{(m)}\right)$,
  where
  $\mathbf g_i^{(m)}$ is the gradient of
  $$l_i^{(m)}(\ttt) = -\sum_{j: \omega_{ij = 1}} \left\{y_{ij}\log f(d_j^{(m)} + \ttt^\top\aaaa_j^{(m)})  + (1-y_{ij})\log(1-f(d_j^{(m)} + \ttt^\top\aaaa_j^{(m)}))\right\}$$
  at $\ttt_i^{(m)}$. $\eta > 0$ is a step size chosen by line search.
  \item[(b)] Given $\ttt_i^{(m+1)}, i = 1, ..., N$ from (a), perform parallel computation for $j = 1, ..., J$. For each item $j$, update
    $(d_j^{(m+1)}, \aaaa_j^{(m+1)}) = \text{Prox}_{C}\left((d_j^{(m)}, \aaaa_j^{(m)}) - \eta \tilde{\mathbf g}_j^{(m)}\right)$,
  where
  $\tilde{\mathbf g}_j^{(m)}$ is the gradient of
  $$\tilde l_j^{(m)}(d, \aaaa) = -\sum_{i: \omega_{ij = 1}} \left\{y_{ij}\log f(d + \aaaa^\top\ttt_i^{(m+1)})  + (1-y_{ij})\log(1-f(d + \aaaa^\top\ttt_i^{(m+1)}))\right\}$$
  at $(d_j^{(m)}, \aaaa_j^{(m)})$. $\eta > 0$ is a step size chosen by line search.
\end{itemize}
Iteratively perform (a) and (b) until convergence.
\item[3] (Output) Output
$\hat \Theta = \Theta^{(M)}$, $\hat A = A^{(M)}$, and $\hat{\mathbf d} = \mathbf d^{(M)}$,
where $M$ is the last iteration number.
\end{enumerate}
\end{algorithm}
The algorithm guarantees the joint likelihood function to increase in each iteration, when the step size $\eta$ in each iteration is properly chosen by line search.
The parallel computing in step 2 of the algorithm is implemented through OpenMP \citep{dagum1998openmp}, which greatly speeds up the computation even on a single machine with multiple cores. The efficiency of this parallel algorithm is further amplified, when running on a computer cluster with many machines. \yc{We also develop a singular value decomposition based algorithm for generating a good starting point for Algorithm~\ref{alg:am}. The details of this algorithm are given in the supplementary material. }



\section{Simulation Study}\label{Sec:simulation}

\subsection{Simulation Study I}

\paragraph{Simulation setting.} In this study, we evaluate the proposed method by Monte Carlo simulation under a variety of settings, listed as follows.
\begin{enumerate}
  \item A growing sequence of number of items is considered: $J = 100, 200, ..., 500$.
  \item Let the number of people $N = \tau J$, where $\tau = 10$ and $25$.
  \item Two choices of $K$ are considered, $K = 3$ and $10$.
\end{enumerate}
This leads to 20 different settings. Under each setting, 100 replications are generated.  For each setting, the true model parameters are generated as follows. We first generate $\ttt_i^0 = (\theta_{i1}^0, ..., \theta_{iK}^0)$ i.i.d. from a $K$-variate truncated normal distribution, for $i = 1, ..., N$. More precisely, the probability density function of $\ttt_i^0$ is given by
$$h_1(\xx) \propto 1_{\{\Vert \xx\Vert \leq 4\sqrt{K}\}} \prod_{k=1}^K \phi(x_k),$$
where $\xx = (x_1, ..., x_K)$ and $\phi(\cdot)$ denotes the probability density function of a standard normal distribution. This truncated normal distribution is very close to a $K$-variate standard normal distribution, since the probability $P(\Vert \XX\Vert \geq 4\sqrt{K})$ is almost 0 when $\XX$ follows a $K$-variate standard normal distribution.
We then generate $d_j^0$ i.i.d. from uniform distribution over the interval $[-2, 2]$, for $j = 1, ..., J$. We finally generate $\aaaa_{j}^0$s, $j =1, ..., J$, so that many of them are sparse.
Specifically, we let
$\mathbf q_j = (q_{j1}, ..., q_{jK})$ be a random vector, satisfying
$$P(\mathbf q_j = \mathbf q) = \frac{1}{2^{K}-1},$$
where $\mathbf q \in \{0, 1\}^K$ and $\mathbf q \neq (0, ..., 0)$.
Also let $\gamma_{jk}$ be i.i.d. uniformly distributed over the interval $[0.5, 2.5]$.  Then we obtain $\aaaa_{j}^0 = (q_{j1}\gamma_{j1}, ..., q_{jK}\gamma_{jK})$.
We obtain $\Theta^*$, $A^*$, and $\mathbf d^*$ by standardizing $\Theta^0 = (\theta_{ik}^0)_{N\times K}$, $A^0 = (a_{jk}^0)_{N\times K}$ and $\mathbf d^0 = (d_1^0, ..., d_J^0)$.

\begin{figure}
\centering
\begin{subfigure}{.5\textwidth}
  \centering
  \includegraphics[width=1\linewidth]{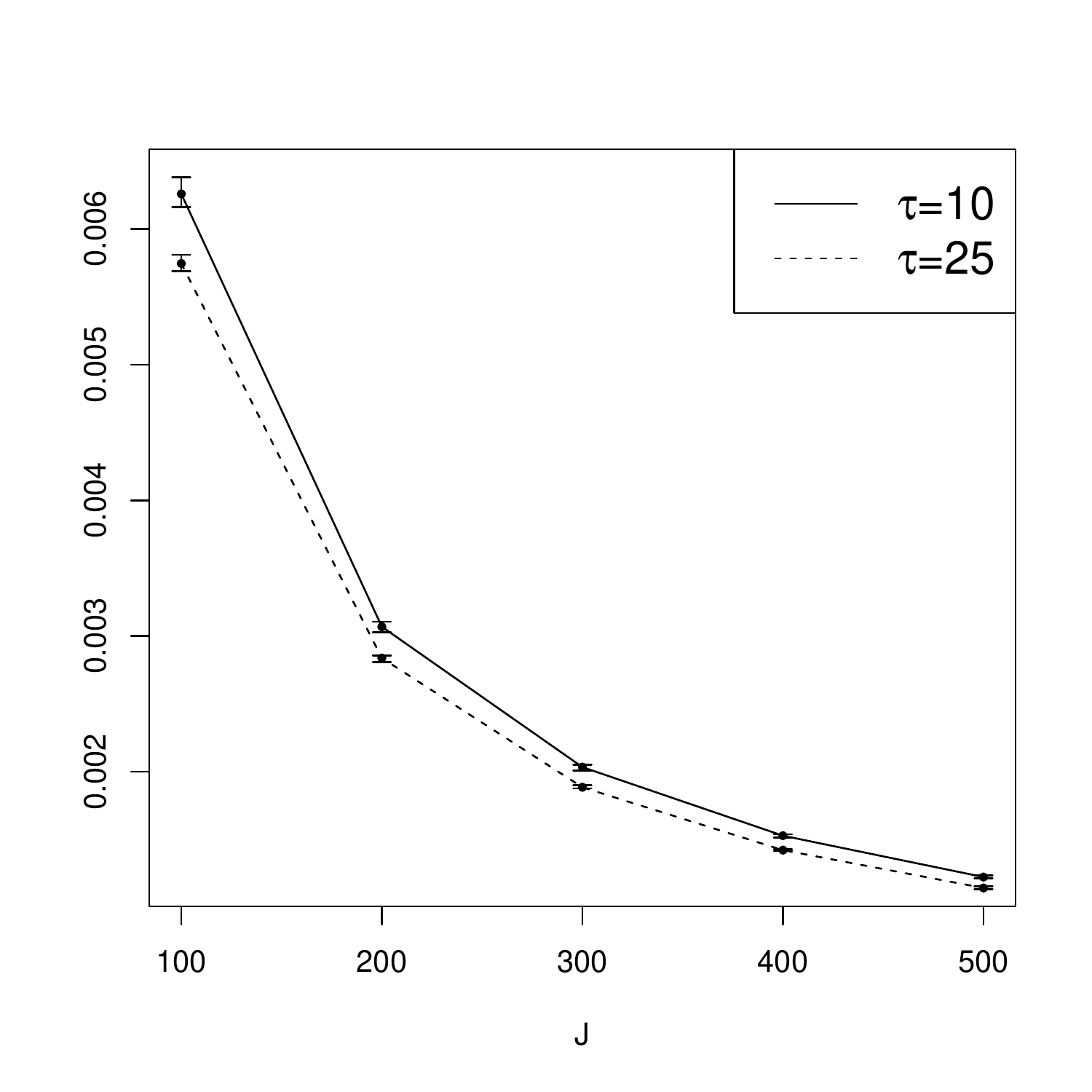}
  \caption{K=3}
\end{subfigure}%
\begin{subfigure}{.5\textwidth}
  \centering
  \includegraphics[width=1\linewidth]{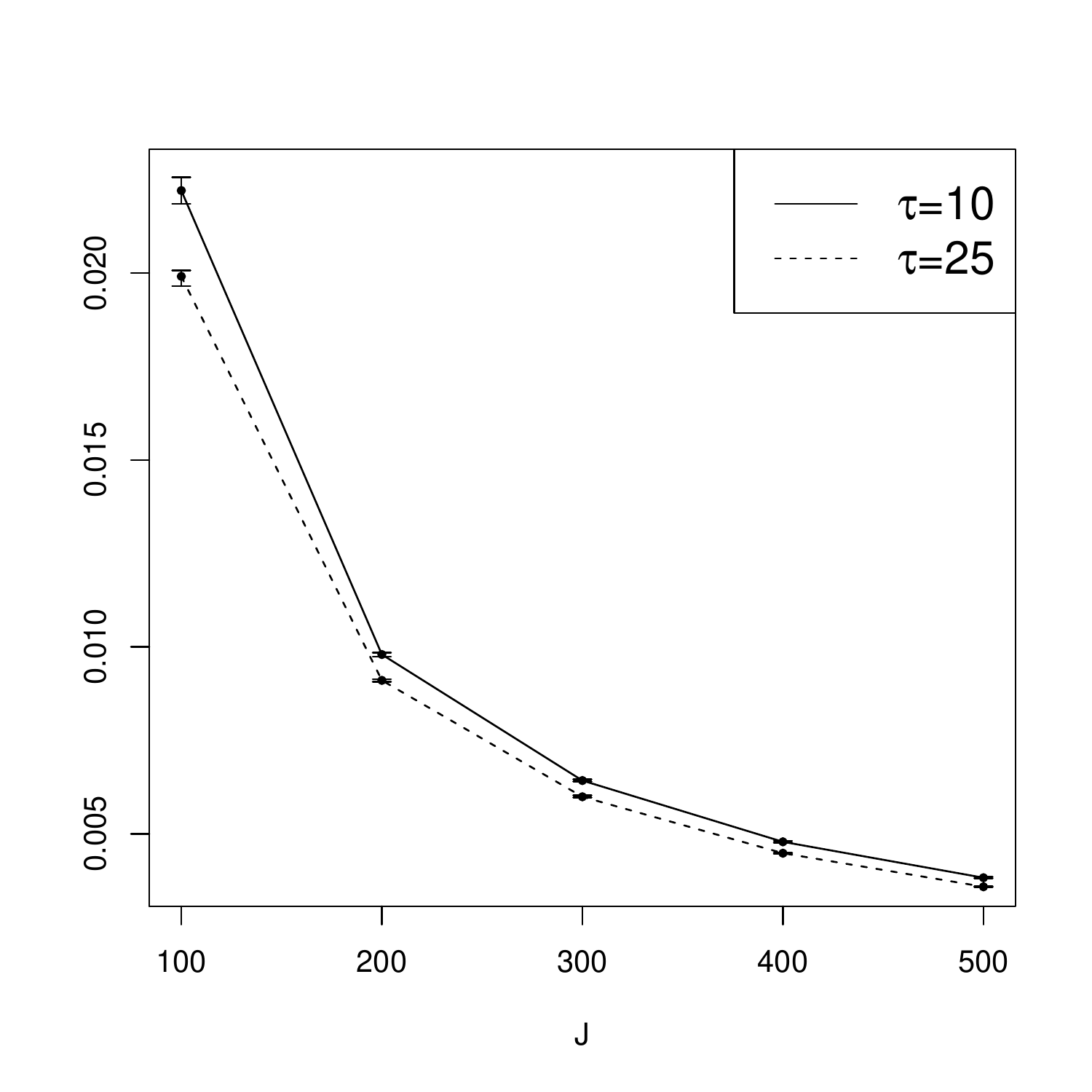}
  \caption{K=10}
\end{subfigure}
  \caption{The scaled Frobenius loss for the recovery of response probabilities when $K=3$ (left panel) and $K=10$ (right panel).}
  \label{fig:pred_error}
\end{figure}

\paragraph{Recovery of response probabilities.} We first show results on the recovery of the response probabilities $f(d_j^* + (\aaaa_j^*)^\top \ttt_i^*)$. Specifically, Figures~\ref{fig:pred_error} shows the value of the scaled Frobenius loss
$$\frac{1}{NJ} \Vert \hat \Theta \hat A^\top + \mathbf 1_N \hat{\mathbf d}^\top  - \Theta^*  (A^*)^\top  - \mathbf 1_N{\mathbf d^*}^\top \Vert^2_F$$
on the $y$-axis versus the number of items $J$ on the $x$-axis, under different settings on the
ratios between $N$ and $J$ and on the latent dimension $K$.
To provide information on the Monte Carlo error, the upper and lower quartiles of the scaled Frobenius loss over 100 replications for each setting are provided.
From these figures, it can be seen that the scaled Frobenius loss decreases as $N$ and $J$ simultaneously increase.

\begin{figure}
\centering
\begin{subfigure}{.5\textwidth}
  \centering
  \includegraphics[width=1\linewidth]{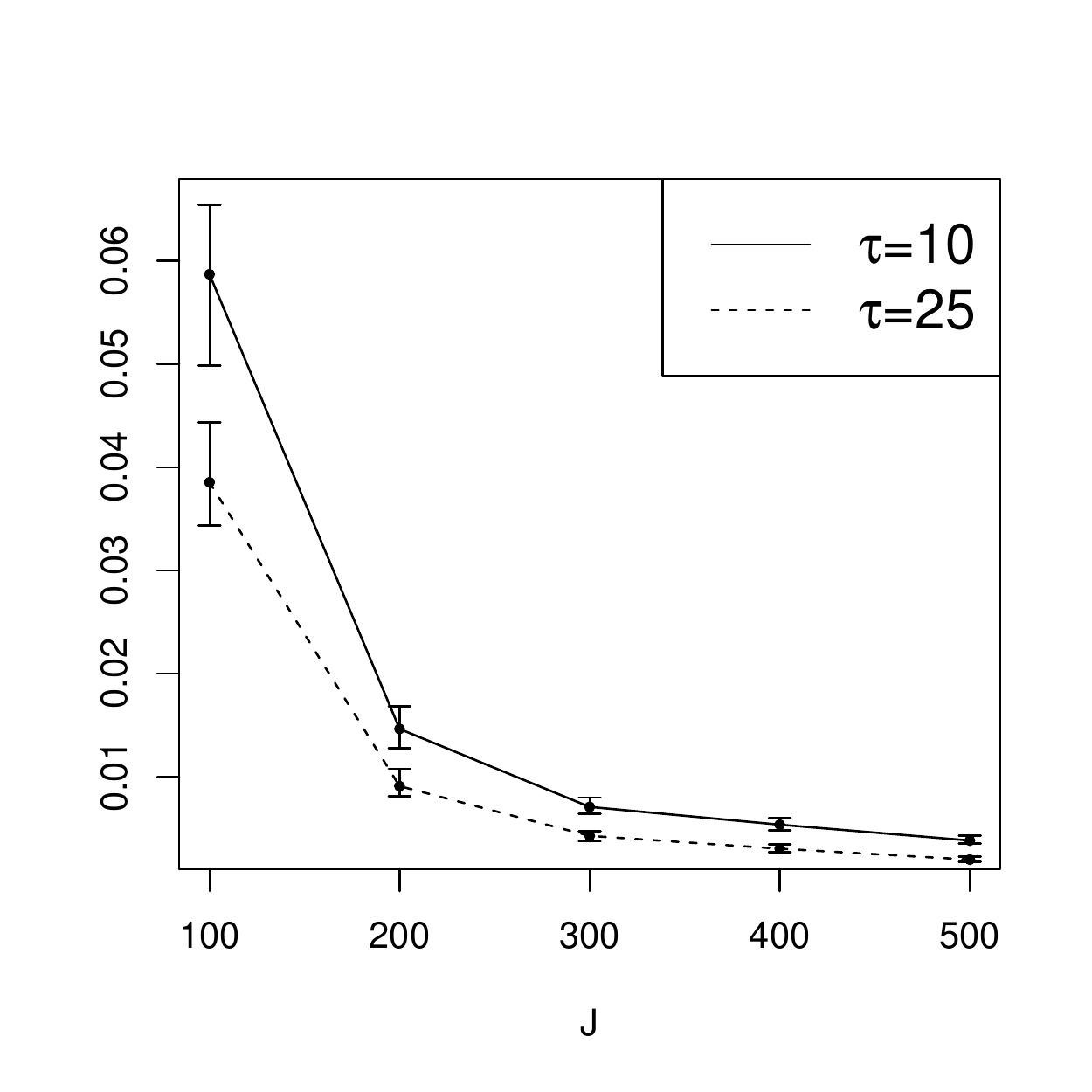}
  \caption{K=3}
\end{subfigure}%
\begin{subfigure}{.5\textwidth}
  \centering
  \includegraphics[width=1\linewidth]{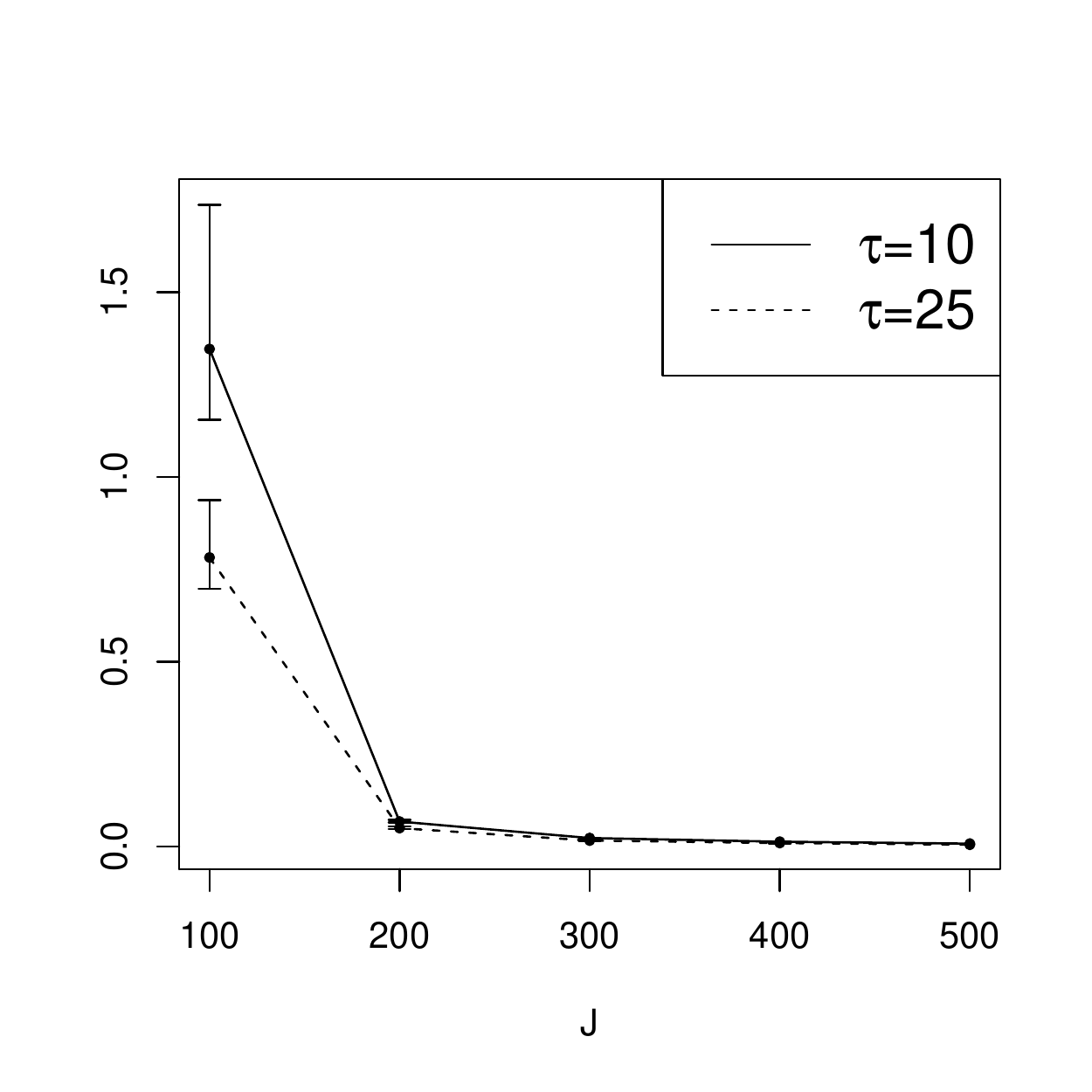}
  \caption{K=10}
\end{subfigure}
  \caption{The scaled Frobenius loss for the recovery of loading matrix up to an orthogonal rotation when $K=3$ (left panel) and $K=10$ (right panel).}
  \label{fig:best_rotation_error}
\end{figure}

\paragraph{Recovery of factor loading matrix up to an orthogonal rotation.} We then show results on the recovery of the
factor loading parameters (up to an orthogonal rotation).  In particular, Figures~\ref{fig:best_rotation_error} shows the
scaled Frobenius loss on the recovery of the loading matrix
$$\min_Q\left\{\frac{1}{JK}\Vert A^* - \tilde A Q\Vert_F^2: Q^\top Q = I_{K\times K} \right\},$$
on the $y$-axis versus the number of items $J$ on the $x$-axis.
These plots are similar to those above for the recovery of response probabilities. Under each setting,  the loss decreases towards zero when $N$ and $J$ simultaneously increase.

\paragraph{Selection of latent dimension by cross validation.} The performance of the cross-validation method for selecting the latent dimension $K$ is evaluated. When the true latent dimension $K = 3$, we consider a candidate set $\{2, 3, 4\}$, and when the  true latent dimension $K = 10$, we choose from $\{9, 10, 11\}$. Five-fold cross-validation is used to choose the latent dimension from the candidate set.
\yc{According to the simulation result, when the true latent dimension $K = 3$, the cross-validation approach always correctly selects $K$. When $K = 10$, 100\% accuracy is achieved except when $J$ is relatively small (68\% for $\tau=10$, $J=100$ and 86\% for $\tau=25$, $J=100$).}

\subsection{Simulation Study II}

\paragraph{Simulation setting.} In this study, we compare the proposed CJMLE with MMLE, where the latter is obtained via an EM algorithm with fixed quadrature points.
We compare under a setting where $K = 2$, since the EM algorithm for MMLE is computationally very intensive when $K$ is larger. We consider a  growing sequence of number of items $J = 100, 200, ..., 500$ and the number of people $N = \tau J$, where $\tau = 10$ and $25$. Each setting is replicated 100 times.

\begin{figure}
\centering
\includegraphics[width=.5\linewidth]{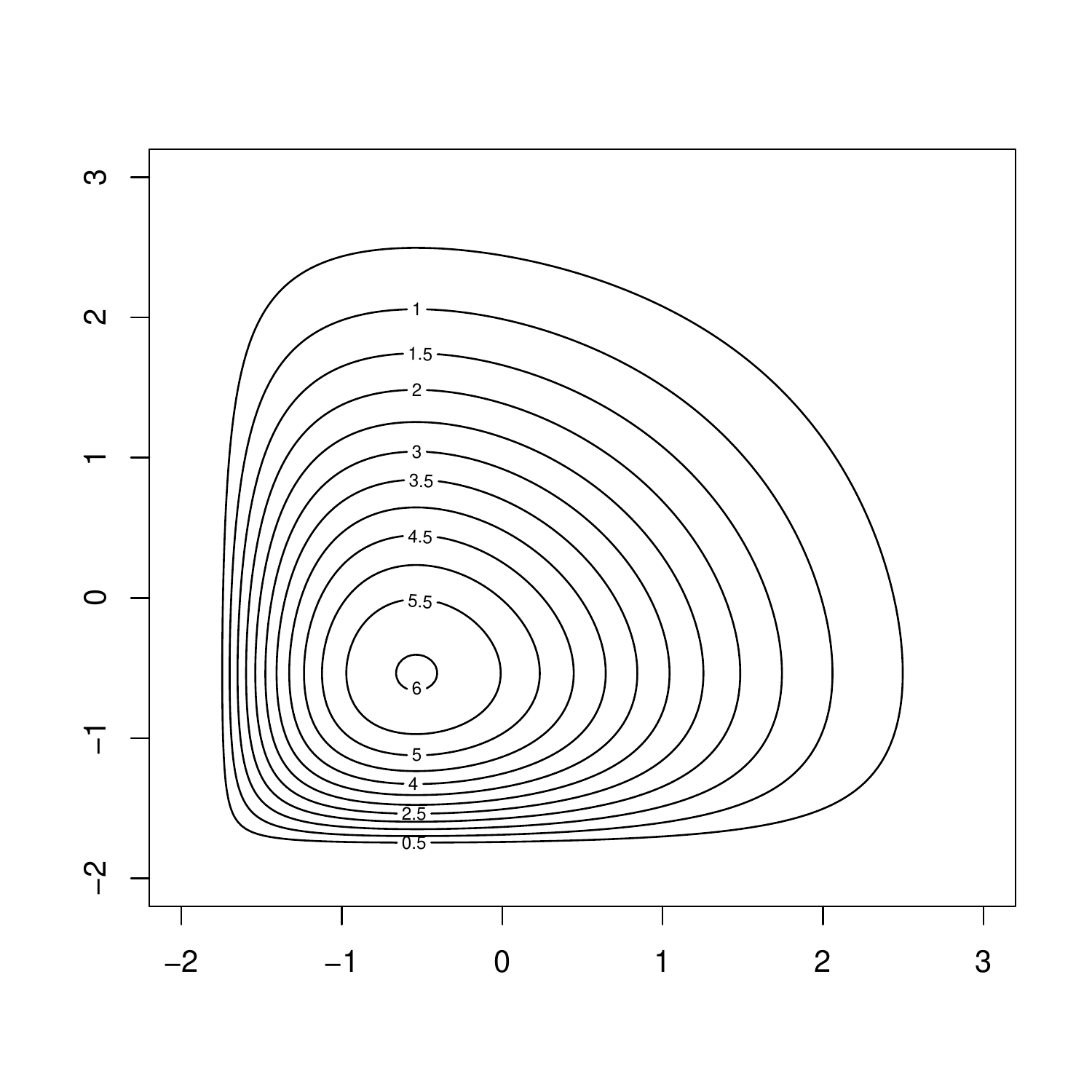}
\caption{The contour plot of the probability density function for $\ttt_i^0$, when $\theta_{i1}^0$ and $\theta_{i1}^0$ are independent and identically distributed, following a scaled and shifted Beta distribution.}
\label{fig:beta25}
\end{figure}

\yc{Two settings are considered for the generation of $\ttt_i^0$. In the first setting, we generate $\ttt_i^0$s i.i.d. from a bivariate standard normal distribution, for $i = 1, ..., N$. In the second setting, we generate $\ttt_i^0$s i.i.d. from a more skewed distribution, by generating $\theta_{i1}^0$ and  $\theta_{i2}^0$ independently from a scaled and shifted Beta distribution. More precisely, we let $$\theta_{ik}^0  = \frac{\zeta_{ik} - \frac{2}{7}}{\sqrt{\frac{5}{196}}}, k = 1, 2, i = 1, ..., N,$$ where $\zeta_{ik}$s are i.i.d. random variables that follow a Beta(2,5) distribution. The scaling and shifting standardizes $\theta_{ik}^0$ to have mean zero and variance one. The distribution of $\ttt_i^0$ is visualized in Figure~\ref{fig:beta25} through a contour plot of its density function.} Given $\ttt_i^0$s,
the item parameters $\aaaa_{j}^0$ and $d_j^{0}$ are generated in the same way as in Study I.
We treat
$\ttt_i^0$, $\aaaa_{j}^0$, and $d_j^{0}$ as the true model parameters and evaluate the two estimation approaches based on (1) the recovery of the response probabilities $f((\aaaa_{j}^0)^\top \ttt_i^0 + d_j^{0})$, (2) the recovery of the factor loading matrix $A^0 = (a_{jk}^0)$ up to an orthogonal rotation, and (3) computation time.

\yc{We point out that under the above simulation setting, the assumption A1 which is required in our theory for the CJMLE is not completely satisfied due to the ways $\ttt_i^0$s are generated. Moreover, in the current implementation of MMLE, a multivariate standard normal distribution is used as the prior for the latent factors. This prior is correctly specified when $\ttt_i^{0}$s are generated from the bivariate standard normal distribution and is misspecified when $\ttt_i^{0}$s are generated from the scaled and shifted Beta distribution.}

The EM algorithm for the MMLE is implemented using the mirt package \citep{chalmers2012mirt} in statistical software R. Specifically, for the numerical integral in the E-step, 31 quadrature points are used for each dimension.
The comparison of computation time is fair, in the sense that both algorithms are implemented in R language with core functions written in C++, given the same starting values, and performed on computers with the same configuration.


\paragraph{Results.}



\begin{figure}
\centering
\begin{subfigure}{.5\textwidth}
  \centering
  \includegraphics[width=1\linewidth]{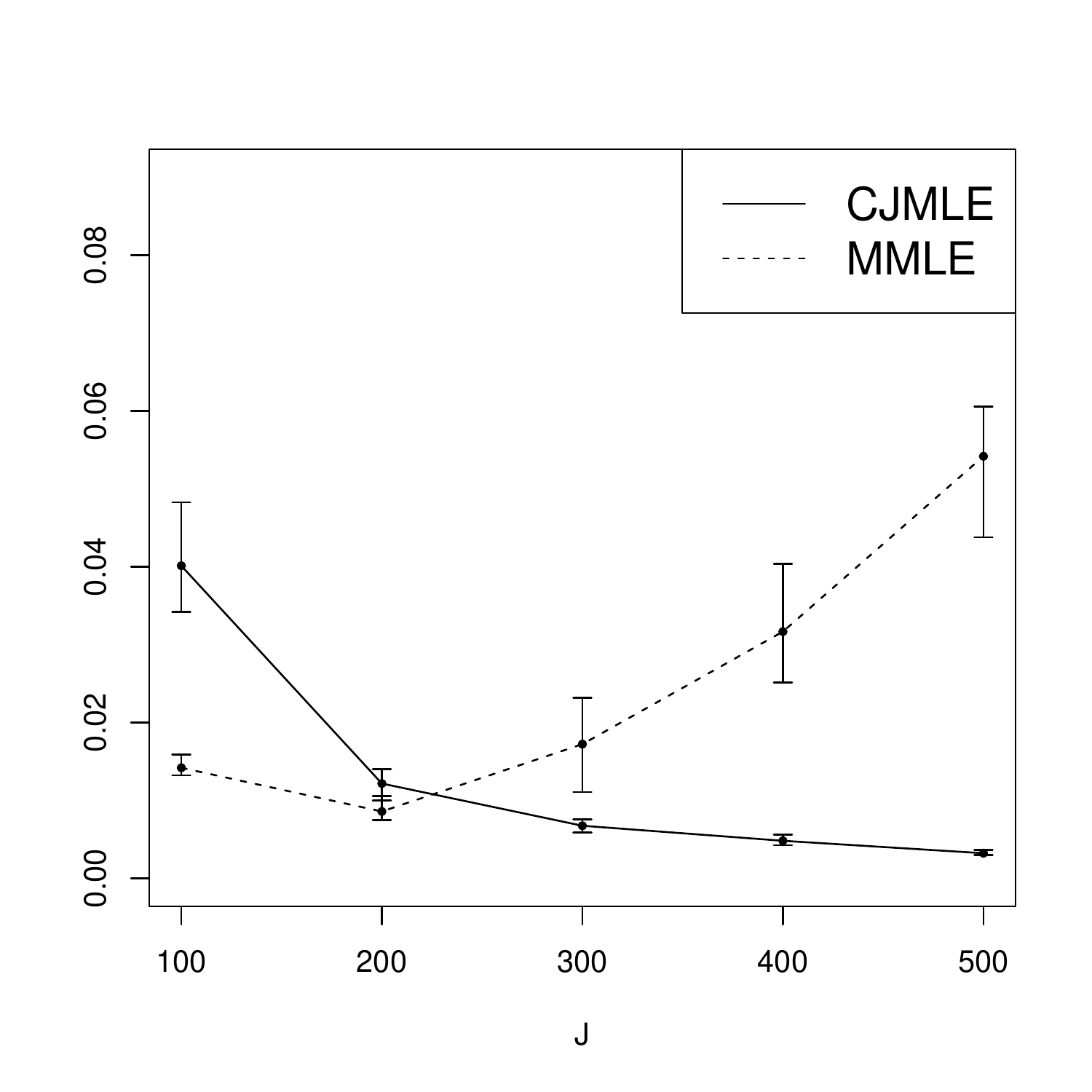}
  \caption{$\tau=10$}
\end{subfigure}%
\begin{subfigure}{.5\textwidth}
  \centering
  \includegraphics[width=1\linewidth]{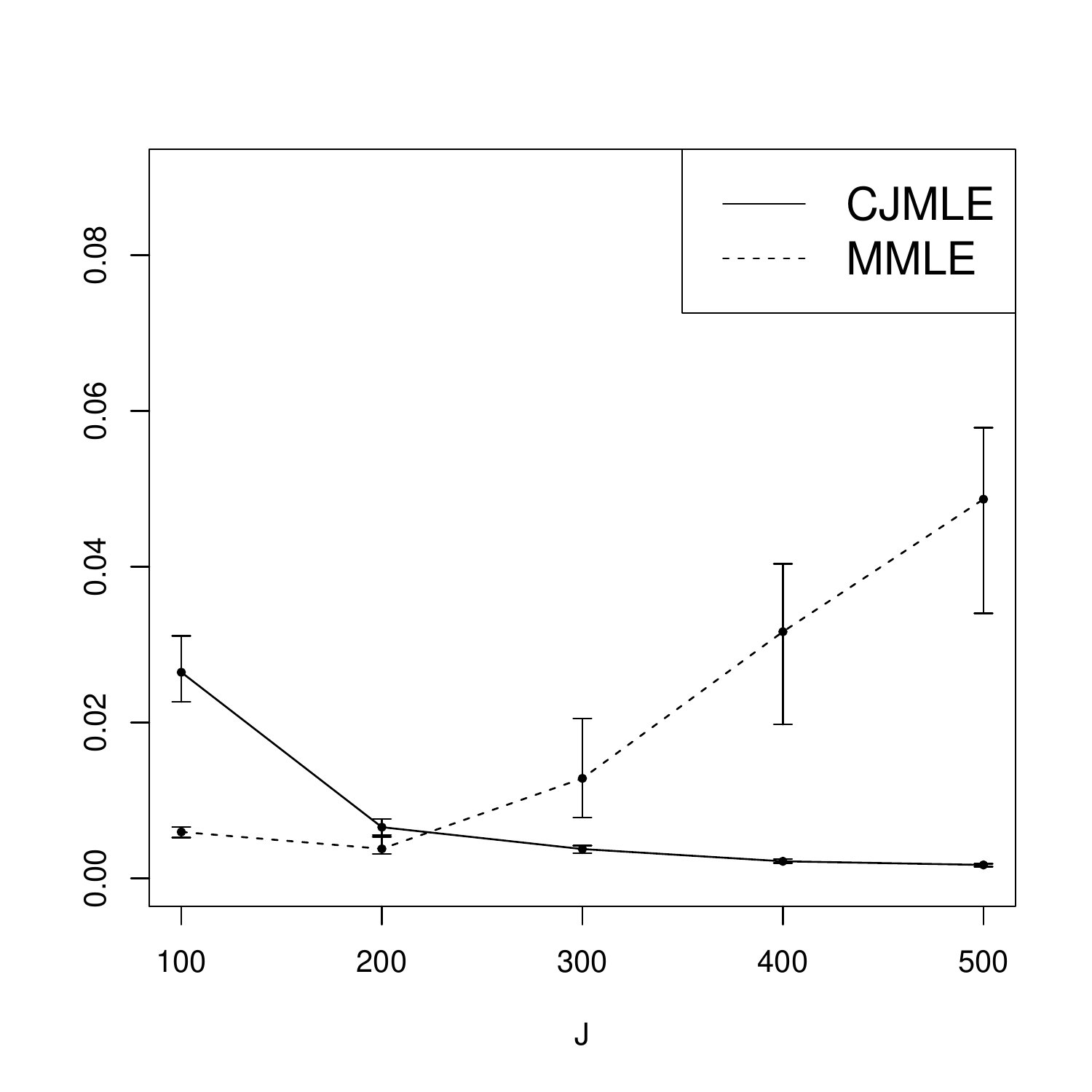}
  \caption{$\tau=25$}
\end{subfigure}
 \caption{Comparison between the CJMLE and MMLE on the recovery of loading matrix up to an orthogonal rotation, when $\ttt_i^0$ follows a standard bivariate normal distribution.}
 \label{fig:loading_comparison}
\end{figure}

\begin{figure}
\centering
\begin{subfigure}{.5\textwidth}
  \centering
  \includegraphics[width=1\linewidth]{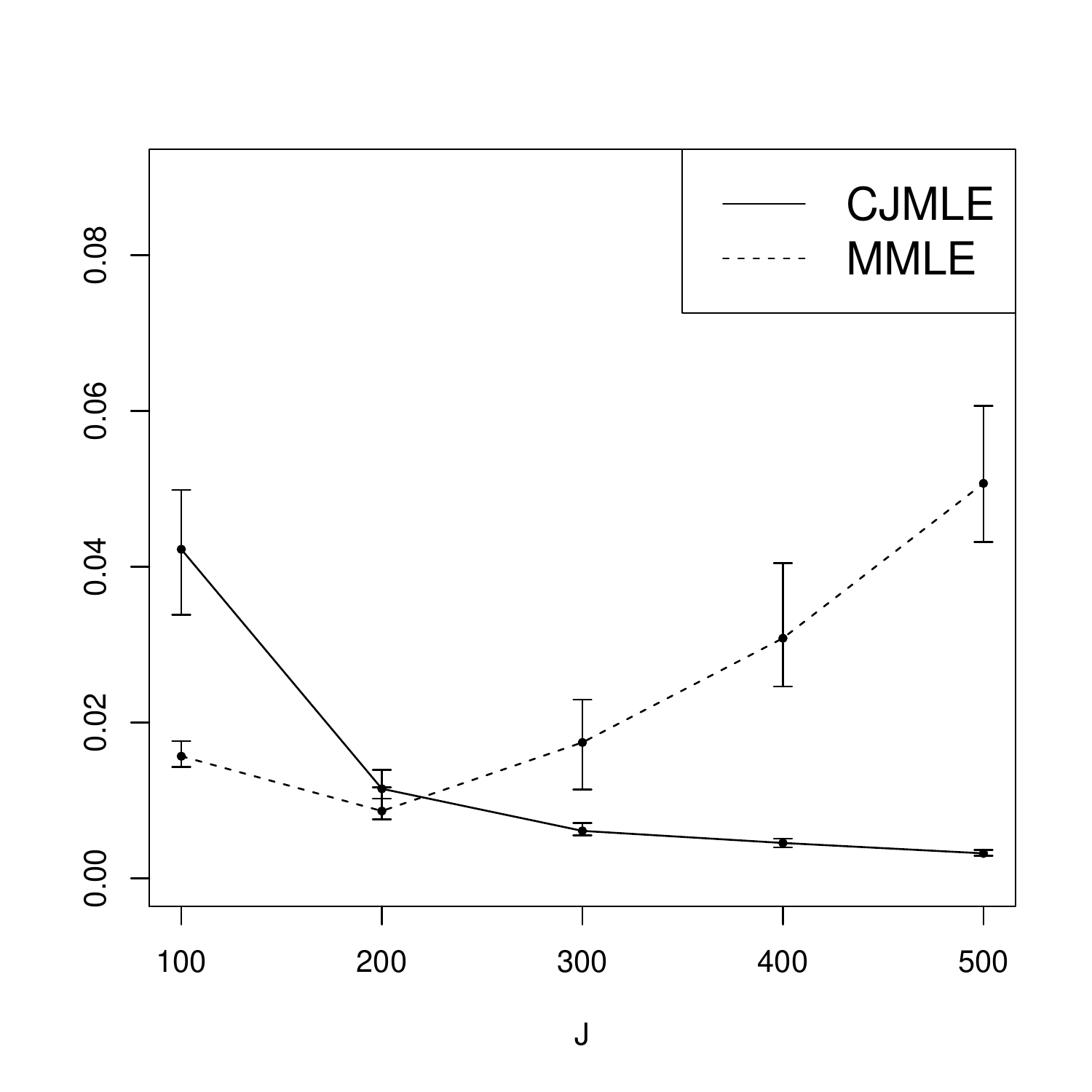}
  \caption{$\tau=10$}
\end{subfigure}%
\begin{subfigure}{.5\textwidth}
  \centering
  \includegraphics[width=1\linewidth]{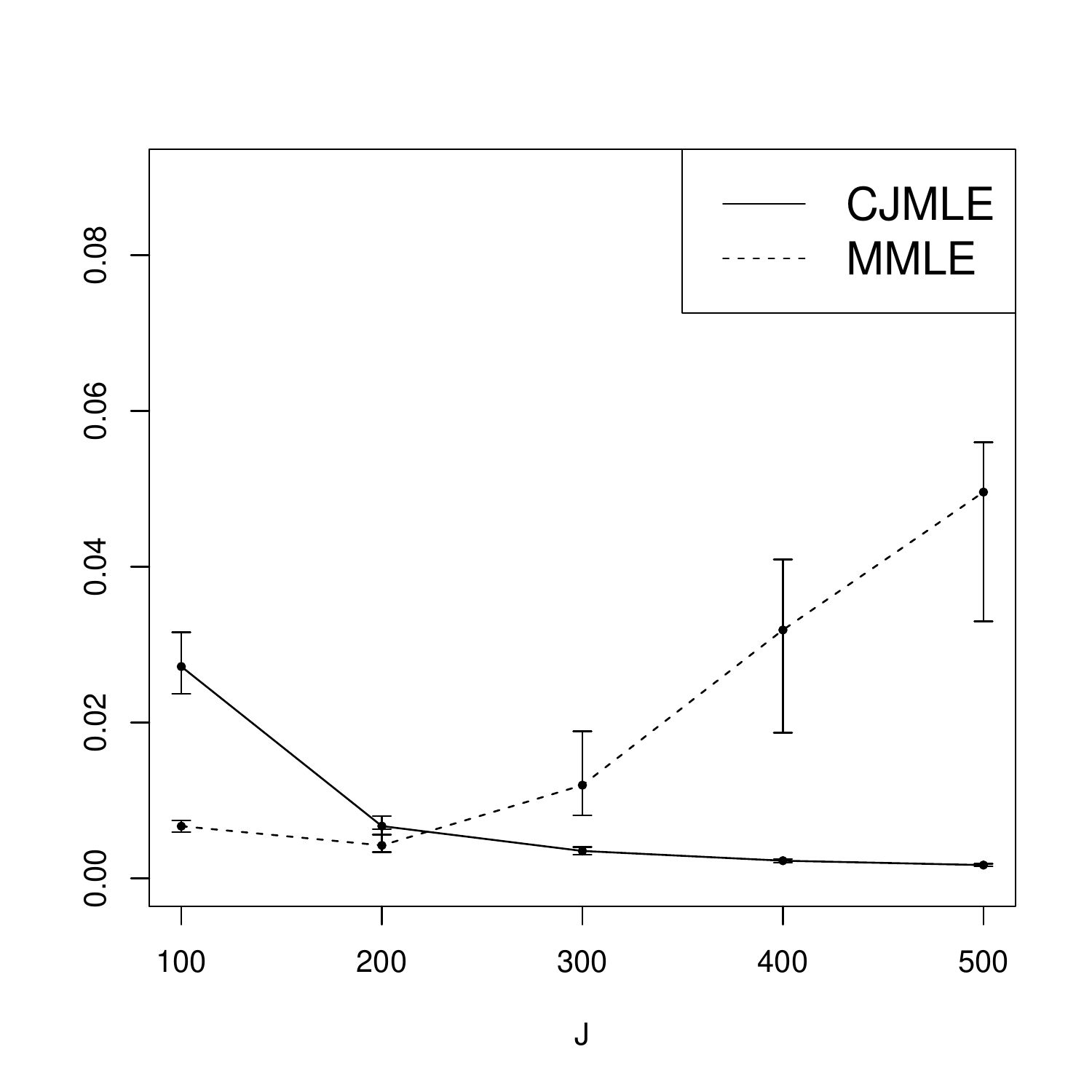}
  \caption{$\tau=25$}
\end{subfigure}
 \caption{Comparison between the CJMLE and MMLE on the recovery of loading matrix up to an orthogonal rotation, when $\ttt_i^0$ is generated based on a Beta distribution.}
 \label{fig:loading_comparison_skew}
\end{figure}

\begin{figure}
\centering
\begin{subfigure}{.5\textwidth}
  \centering
  \includegraphics[width=1\linewidth]{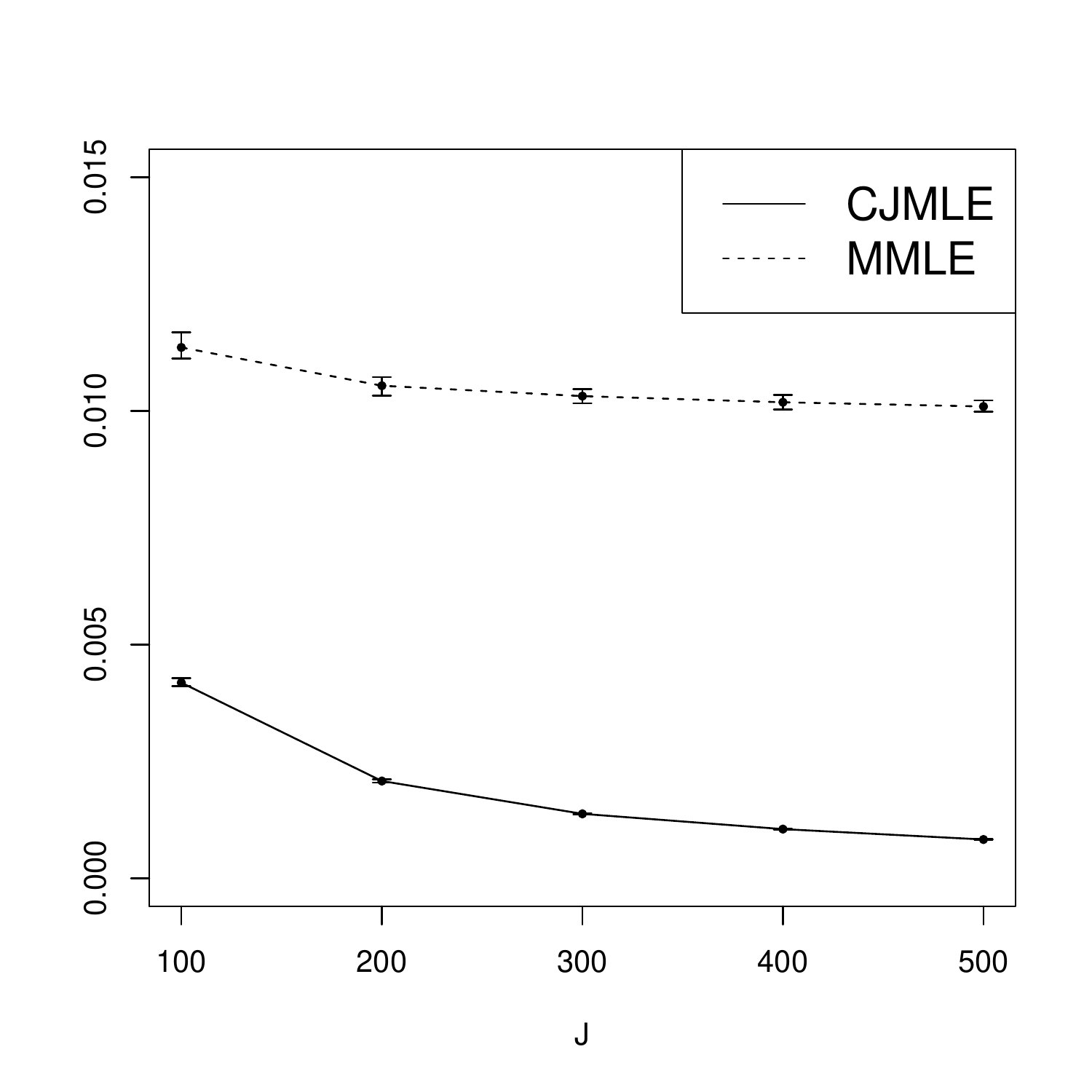}
  \caption{$\tau=10$}
\end{subfigure}%
\begin{subfigure}{.5\textwidth}
  \centering
  \includegraphics[width=1\linewidth]{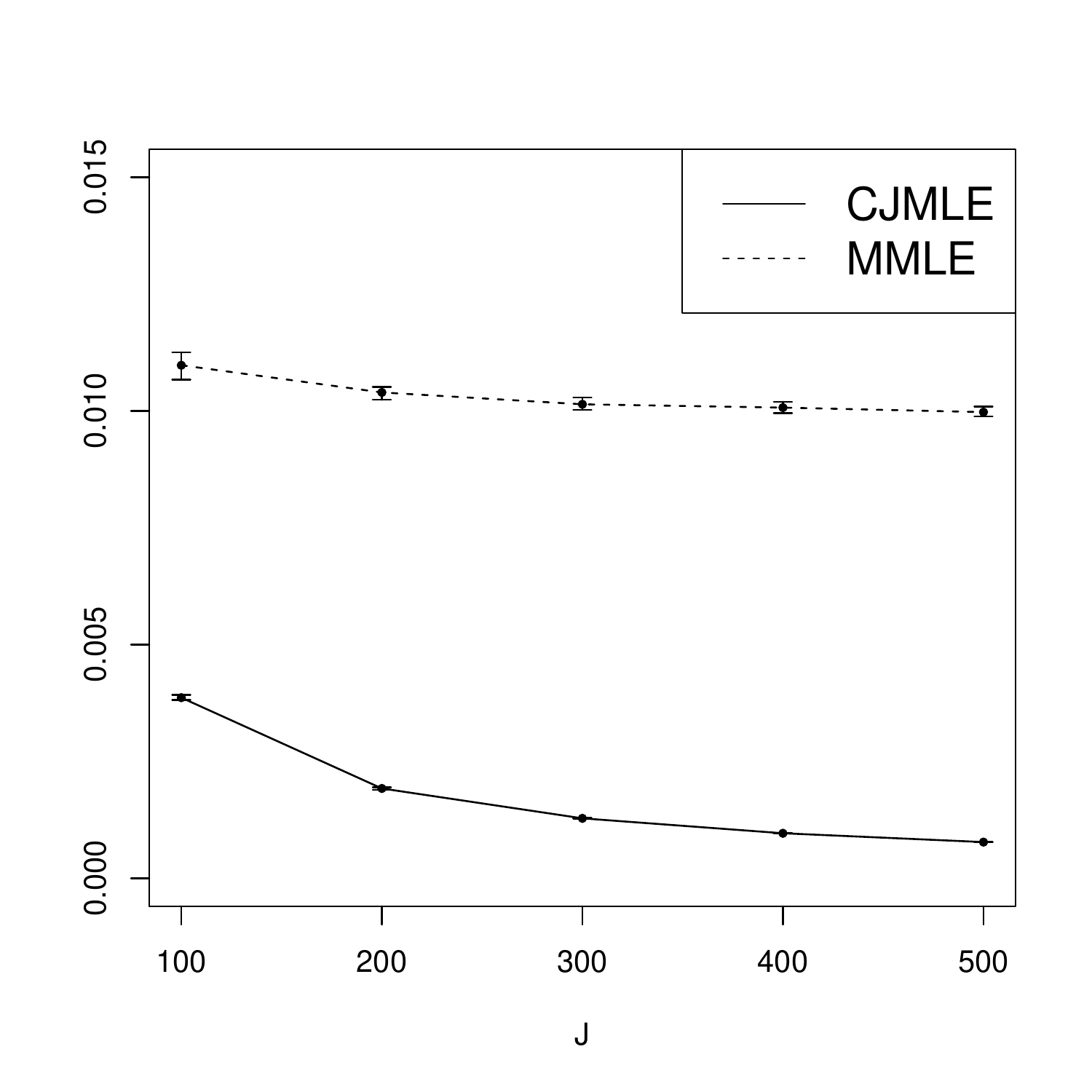}
  \caption{$\tau=25$}
\end{subfigure}
 \caption{Comparison between the CJMLE and MMLE on the recovery of the item response probabilities, when $\ttt_i^0$ follows a standard bivariate normal distribution.}
 \label{fig:pred_error_comparison}
\end{figure}

\begin{figure}
\centering
\begin{subfigure}{.5\textwidth}
  \centering
  \includegraphics[width=1\linewidth]{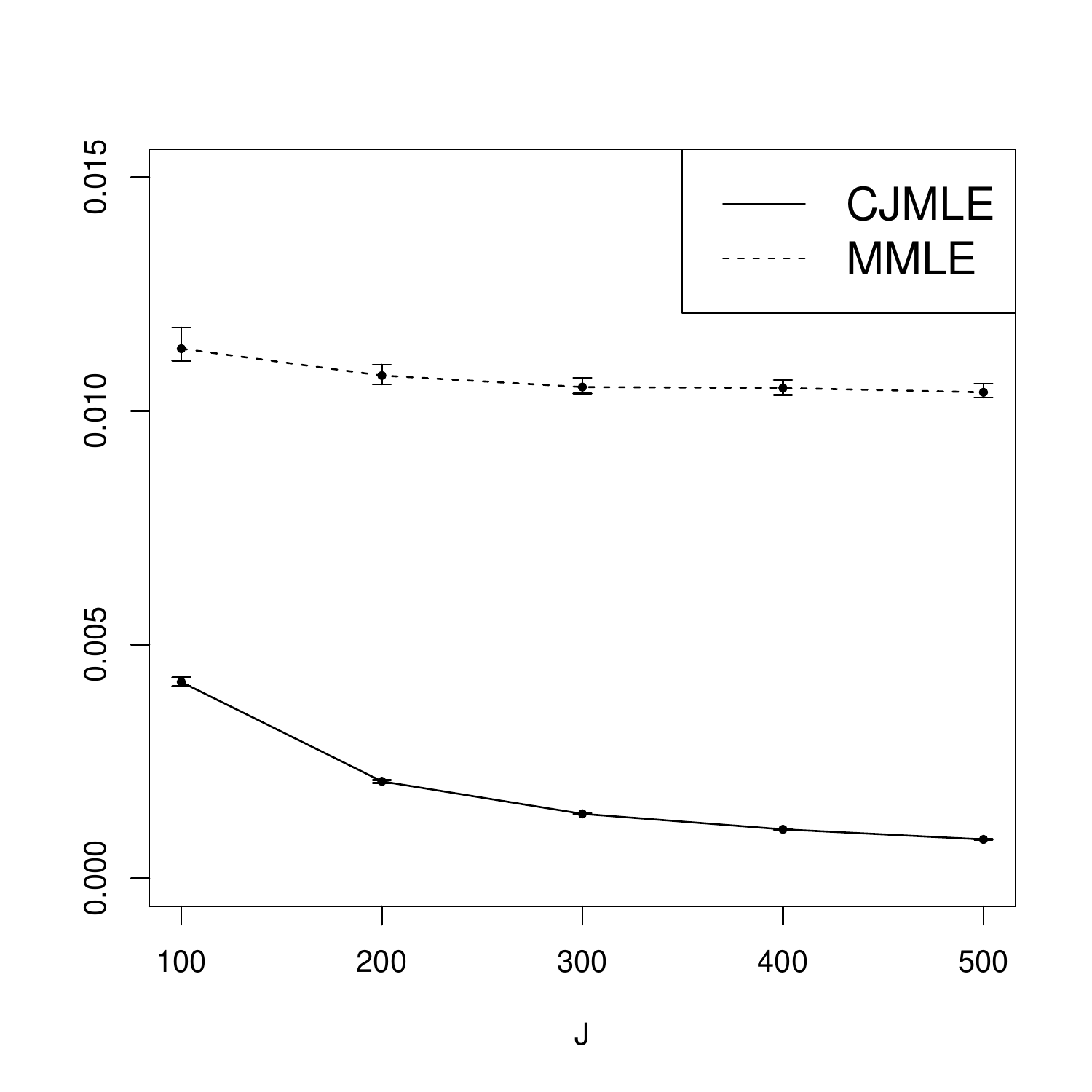}
  \caption{$\tau=10$}
\end{subfigure}%
\begin{subfigure}{.5\textwidth}
  \centering
  \includegraphics[width=1\linewidth]{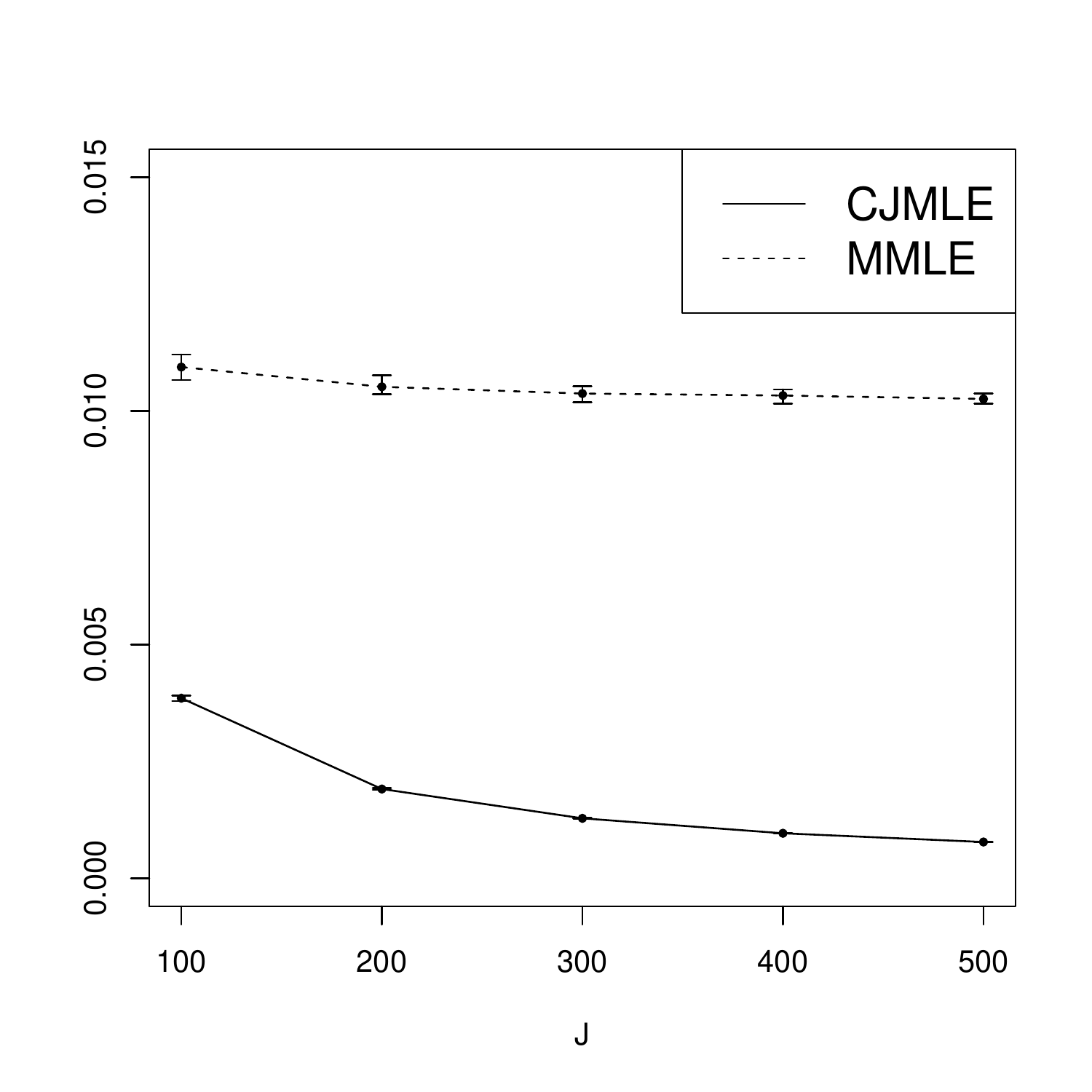}
  \caption{$\tau=25$}
\end{subfigure}
 \caption{Comparison between the CJMLE and MMLE on the recovery of the item response probabilities, when $\ttt_i^0$ is generated based on a Beta distribution.}
 \label{fig:pred_error_comparison_skew}
\end{figure}

\yc{The results are given in Figures~\ref{fig:loading_comparison} through \ref{fig:pred_error_comparison_skew} and Tables~\ref{tab:speed_comparison} and \ref{tab:speed_comparison_skew},
where the results are similar under both settings for $\ttt_i^0$. In terms of the recovery of the loading matrix up to an orthogonal rotation, as shown in Figures~\ref{fig:loading_comparison} and \ref{fig:loading_comparison_skew}, the MMLE performs better when $N$ and $J$ are small and the CJMLE outperforms the MMLE when both $N$ and $J$ are sufficiently large, regardless of the ways $\ttt_i^0$s are generated. It is also observed that the scaled Frobenius loss keeps decreasing for the CJMLE when $N$ and $J$ grow simultaneously, which is not the case for the MMLE. For the MMLE,
even when the prior distribution is correctly specified for the latent traits, the scaled Frobenius loss for the recovery of the loading matrix first decreases and then increases when $N$ and $J$ simultaneously increase. This is possibly due to the approximation error brought by the fixed quadrature points and is worth future investigation from a theoretical perspective. In terms of the recovery of the item response probabilities based on the scaled Frobenius loss, which is presented in Figures~\ref{fig:pred_error_comparison} and \ref{fig:pred_error_comparison_skew}, the CJMLE always outperforms the MMLE throughout all the settings. Finally, according to Tables~\ref{tab:speed_comparison} and \ref{tab:speed_comparison_skew}, the CJMLE is substantially faster than the EM algorithm for MMLE. For example, when $J = 500, N = 5000$, and $\ttt_i^0$ follows a bivariate standard normal distribution,
the median computation time for
the CJMLE is 80 seconds, while that for the MMLE via the EM algorithm is more than 2,000 seconds. }


%

\begin{table}
  \centering
  \begin{tabular}{r|ccccc}
  \hline
  $\tau=10$ &J=100&J=200&J=300&J=400&J=500\\
  \hline
     (25\% quantile) & 1.5 & 6.2 & 16.6 & 35.7 & 78.8\\
CJMLE (50\% quantile) &1.5 & 7.5 & 16.7 & 36.1 & 80.2\\
     (75\% quantile) &1.5 & 7.6 & 20.2 & 36.5 & 80.9\\
\hline
(25\% quantile) &44.9 & 157.1 & 354.9 &  771.6 & 1599.5\\
MMLE (50\% quantile) &78.0 & 333.4 & 500.0 & 1079.0 & 2008.5\\
(75\% quantile) &93.6 & 459.8 & 745.5 & 1637.8 & 2932.5\\
\hline
\hline
  $\tau=25$&J=100&J=200&J=300&J=400&J=500\\
    \hline
(25\% quantile) & 4.0 & 16.2 & 43.6 & 95.0 & 198.6 \\
CJMLE (50\% quantile) &4.0 & 16.3 & 43.8 & 95.9 & 211.0 \\
 (75\% quantile) &4.0 & 16.4 & 53.2 & 96.4 & 245.0 \\
\hline
(25\% quantile) & 75.5 & 511.1 & 1095.4 & 1741.2 & 2799.4 \\
MMLE (50\% quantile) &145.9 & 741.8 & 2227.8 & 2901.5 & 3742.4 \\
(75\% quantile) &186.1 & 898.0 & 3038.1 & 4785.2 & 6387.0 \\
\hline
\end{tabular}
\caption{Speed comparison (in seconds) between CJMLE and MMLE measured in seconds on a single Intel$^\circledR$ E5-2650v4 core, when $\ttt_i^0$ follows a standard bivariate normal distribution.}\label{tab:speed_comparison}
\end{table}

\begin{table}
  \centering
  \begin{tabular}{r|ccccc}
  \hline
  $\tau=10$ &J=100&J=200&J=300&J=400&J=500\\
  \hline
     (25\% quantile) & 1.4 & 5.4 & 14.1 & 30.9 & 64.7\\
CJMLE (50\% quantile)& 1.4 & 5.5 & 14.2 & 34.5 & 70.8\\
     (75\% quantile) & 1.4 & 6.7 & 17.6 & 35.3 & 72.7\\
\hline
(25\% quantile)       & 59.1 & 154.5 & 344.3 & 783.3 & 1583.1\\
MMLE (50\% quantile)  & 81.9 & 289.3 & 522.3 & 987.1 & 2059.3\\
(75\% quantile)       & 102.9 & 477.2 & 782.8 & 1455.0 & 2958.3\\
\hline
\hline
  $\tau=25$&J=100&J=200&J=300&J=400&J=500\\
    \hline
(25\% quantile)       & 3.7 & 14.4 & 37.5 & 85.4 & 164.0 \\
CJMLE (50\% quantile) & 3.7 & 14.6 & 37.6 & 87.3 & 180.5 \\
 (75\% quantile)      & 3.7 & 17.9 & 37.8 & 89.1 & 189.1 \\
\hline
(25\% quantile)      & 81.2 & 363.1 & 1262.1 & 1624.4 & 2651.7\\
MMLE (50\% quantile) & 145.7 & 685.4 & 2222.1 & 2259.0 & 3390.7\\
(75\% quantile)      & 189.5 & 850.6 & 3033.7 & 4369.4 & 7164.7\\
\hline
\end{tabular}
\caption{Speed comparison (in seconds) between CJMLE and MMLE measured in seconds on a single Intel$^\circledR$ E5-2650v4 core, when $\ttt_i^0$ is generated based on a Beta distribution.}\label{tab:speed_comparison_skew}
\end{table}

\begin{figure}
\centering
\begin{subfigure}{.5\textwidth}
  \centering
  \includegraphics[width=1\linewidth]{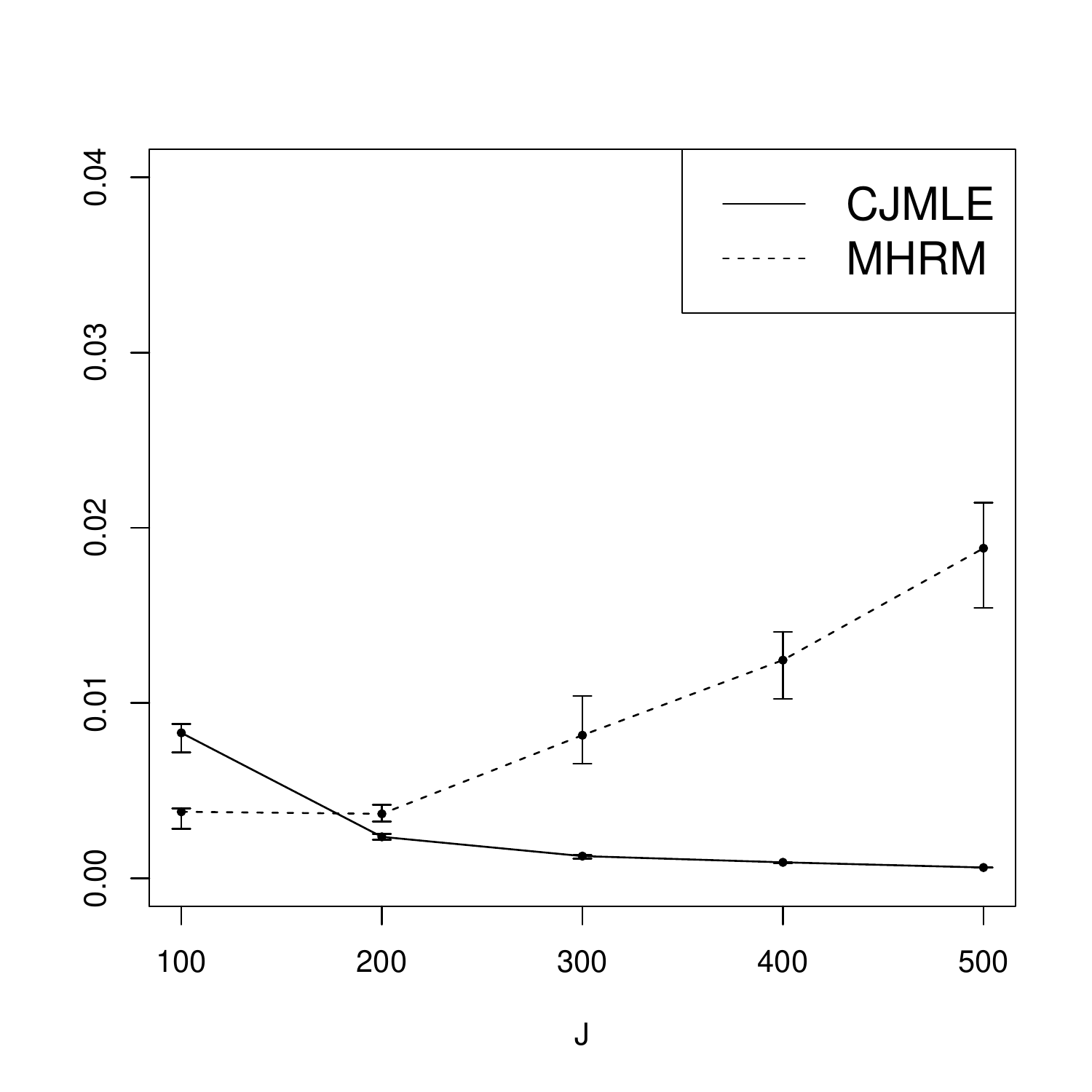}
  \caption{Recovery of loading matrix.}
\end{subfigure}%
\begin{subfigure}{.5\textwidth}
  \centering
  \includegraphics[width=1\linewidth]{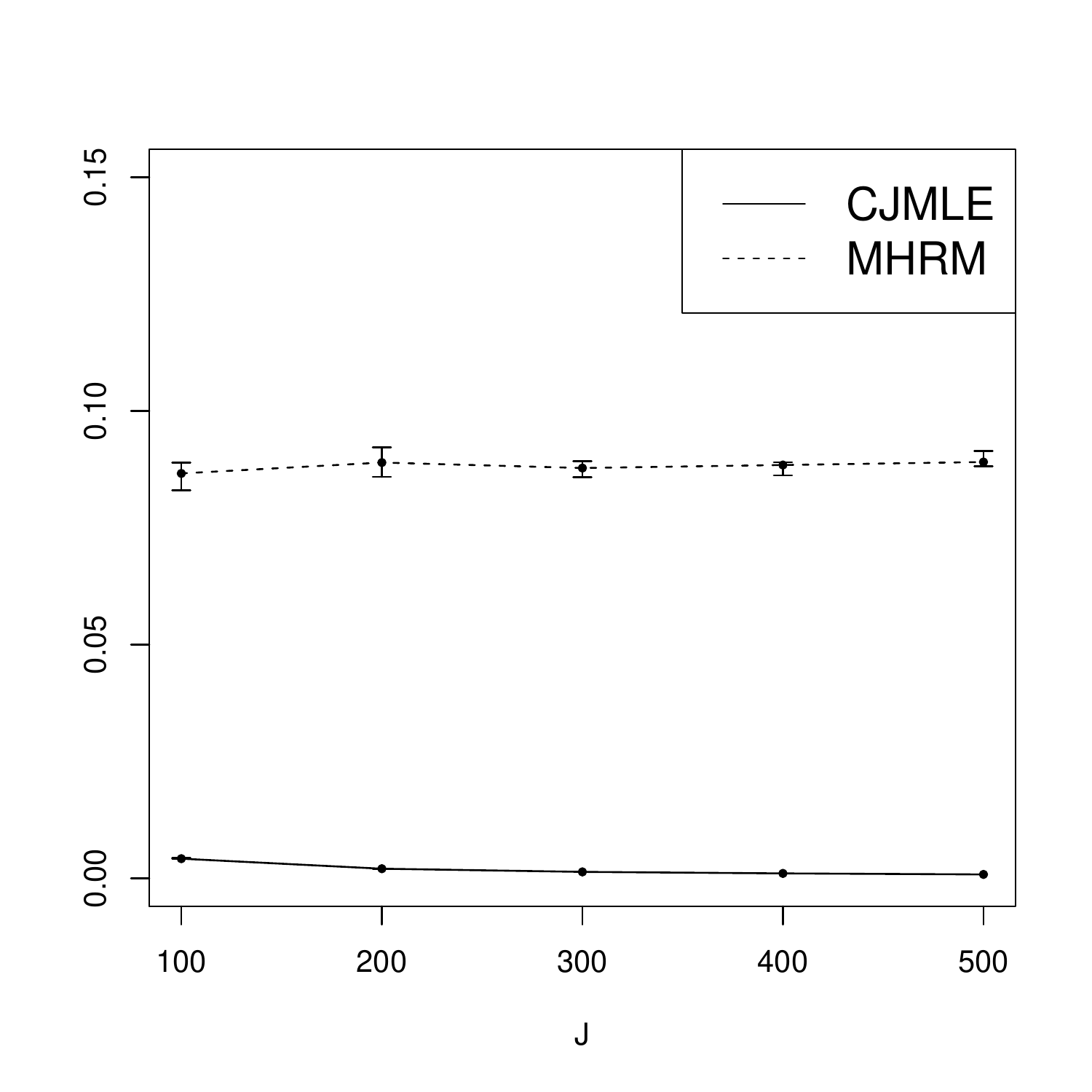}
  \caption{Recovery of item response probabilities.}
\end{subfigure}
 \caption{Comparison between the CJMLE and MHRM algorithms.}
 \label{fig:comparison_flexmirt}
\end{figure}

\yc{
\subsection{Simulation Study III}
We further compare the proposed CJMLE algorithm with
a Metropolis-Hastings Robbins-Monro (MHRM) algorithm \citep{cai2010high, cai2010metropolis}, which is one of the state-of-the-art algorithms for high-dimensional item factor analysis. This algorithm is implemented in IRT software
\textbf{flexMIRT}$^\circledR$.

\paragraph{Simulation setting.} We compare under a setting where $K = 10$, the number of items $J=100, 200, \cdots, 500,$ and the number of people $N=10 J$. We generate $\ttt_i^0$s i.i.d. from a bivariate standard normal distribution, for $i = 1, ..., N$. The item parameters $\aaaa_{j}^0$ and $d_j^{0}$ are generated in the same way as in Study I. Each setting is replicated 10 times\footnote{The small number of replications is due to the constraint that
\textbf{flexMIRT}$^\circledR$ needs to be run on a local Windows$^\circledR$ machine.}.


\paragraph{Results.} The two algorithms are compared under the same criteria as in Study~II. The results are shown in Figure~\ref{fig:comparison_flexmirt} and Table~\ref{tab:speed_comparison_flexmirt}. According to these results, under the current setting,  the CJMLE is not only much faster than the MHRM method, but also more accurate in terms of the recovery of factor loading parameters when $J\geq 200$ (panel (a) of Figure~\ref{fig:comparison_flexmirt}) and in terms of the recovery of
item response probabilities (panel (b) of Figure~\ref{fig:comparison_flexmirt}). It is noticed that similar to the result of Study II,  the scaled Frobenius loss for the recovery of the loading matrix keeps increasing when $N$ and $J$ simultaneously increase. It may be due to that the default stopping criterion in \textbf{flexMIRT}$^\circledR$ for the MHRM algorithm
does not adapt well to the simultaneous growth of $N$ and $J$.
}




\begin{table}
  \centering
  \begin{tabular}{r|ccccc}
  \hline
  $\tau=10$ &J=100&J=200&J=300&J=400&J=500\\
  \hline
     (25\% quantile)  & 1.8 &  8.6 & 29.0 & 68.3 & 137.9\\
CJMLE (50\% quantile) & 1.9 & 10.0 & 29.6 & 68.7 & 138.4\\
     (75\% quantile)  & 1.9 & 10.0 & 34.7 & 69.0 & 139.4\\
\hline
(25\% quantile)      & 142.8 & 478.7 & 850.4 & 1644.8 & 1997.6\\
MHRM (50\% quantile) & 163.2 & 574.4 & 1077.6 & 2147.0 & 2573.4\\
(75\% quantile)      & 189.7 & 609.7 & 1157.8 & 2403.1 & 2894.3\\
\hline
\end{tabular}
\caption{Speed comparison (in seconds) between the CJMLE and the MR-HM measured in seconds on a
 single Intel$^\circledR$ core (Xeon$^\circledR$ CPU @2.20 GHz; RAM 3.75 GB).}\label{tab:speed_comparison_flexmirt}
\end{table}

\yc{\section{Real Data Analysis}\label{Sec:real}}

We illustrate the use of the proposed method on
the female UK normative sample data for the EPQ-R \citep{eysenck1985revised}. The dataset contains
the responses to 79 dichotomous items from 824 people.
Among these items, items 1-32, 33-55, and 56-79 consist of the Psychoticism (items 1-32), Extraversion (items 33-55) and Neuroticism (items 56-79) scales, respectively, which are designed to measure the corresponding personality traits.
The data have been pre-processed so that the negatively worded items are reversely scored.
We analyze the dataset in an exploratory manner and then compare the results with the design of the items.

\begin{figure}
\centering
\includegraphics[width=0.5\linewidth]{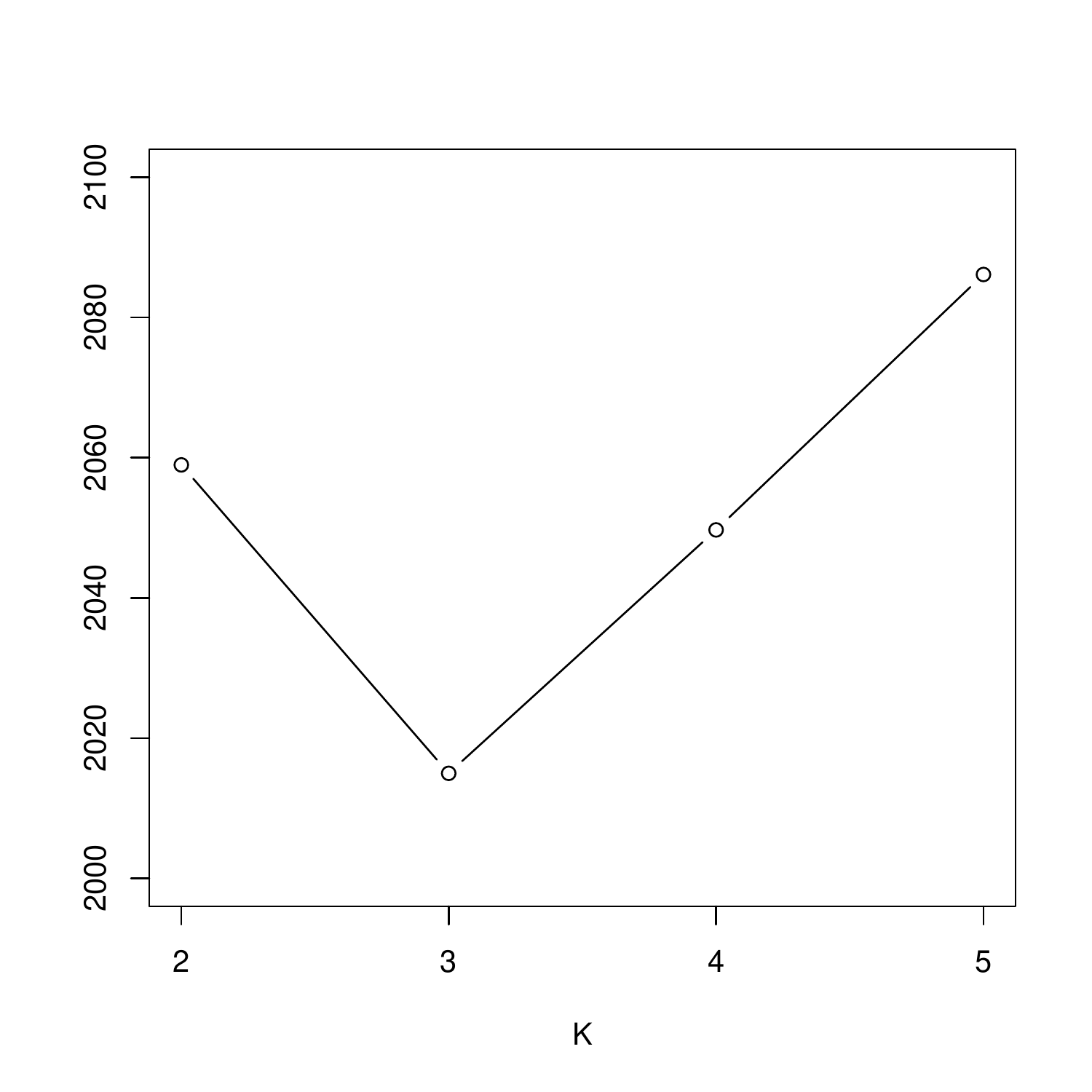}
 \caption{Cross validation errors for $K = 2, 3, 4, 5$.}
 \label{fig:choose_k}
\end{figure}

\paragraph{Selection of number of factors.} We first select the latent dimension $K$ using a five-fold cross-validation method, as described in Section~\ref{subsec:numFac}.
The result is given in Figure~\ref{fig:choose_k}, where the smallest cross-validation error is achieved when $K=3$.
This result is consistent with the design of the EPQ-R.
In what follows, we report the estimated parameters under the three-factor model.

\paragraph{Three-factor model result.} To anchor the latent factors, we apply an analytic rotation method, the Geomin rotation \citep[see e.g.,][]{yates1988multivariate}, to the obtained three-factor solution. Geomin is an oblique rotation method that aims at finding a simple pattern of factor loadings without requiring the factors to be orthogonal to each other.

\begin{figure}
\centering
\includegraphics[width=1\linewidth]{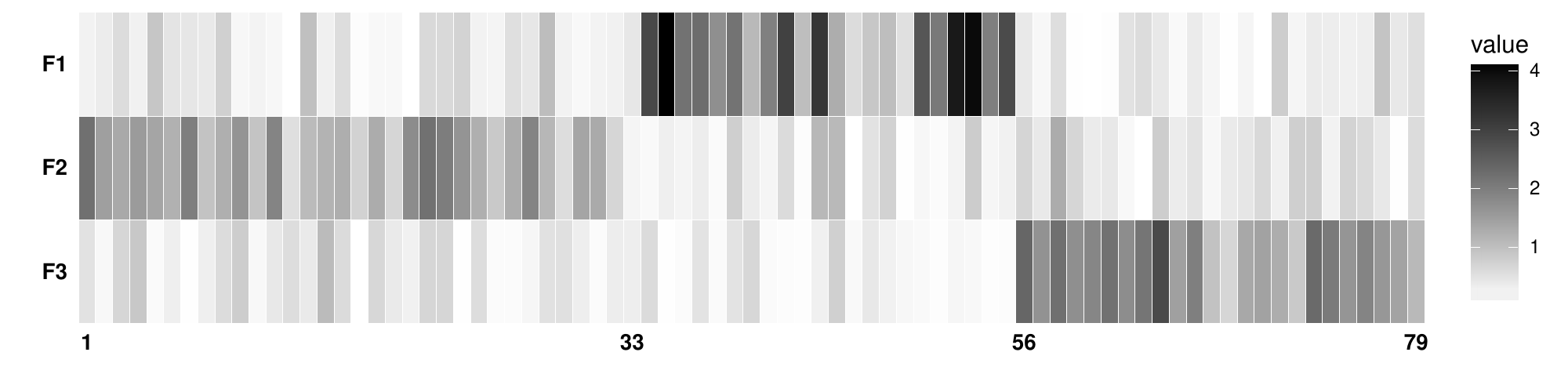}
 \caption{Fitting a three-factor model to the EPQ-R data:  Heat map of the estimated loading matrix in absolute value under Geomin rotation.}
 \label{fig:loading_heat}
\end{figure}

In Figure~\ref{fig:loading_heat}, we present a heat map of the estimated factor loading matrix in absolute values. As we can see, items in the E, P, and N scales tend to have large absolute loadings on the three estimated factors, respectively. We list the top five items with the highest absolute loadings on each factor in Table~\ref{tab:3factor}. These items are all from the corresponding scales and are quite representative of the scales that they belong to.
The correspondence between the recovered factors and the Eysenck's three personality traits
is further confirmed by the high correlations between the estimated person parameters (after rotation) and the corresponding total scores on the three scales, as given in Table~\ref{tab:score_cor}.

\begin{figure}
\centering
\begin{subfigure}[b]{1\textwidth}
\includegraphics[width=1\linewidth]{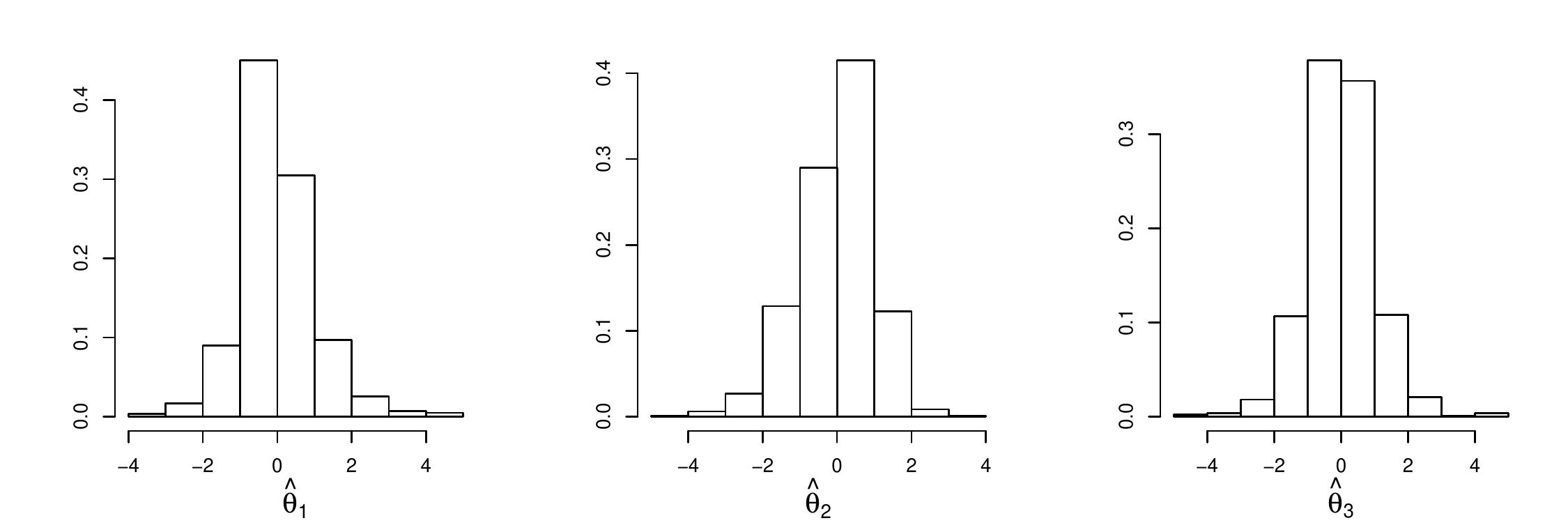}
 \caption{Histograms of the estimated person parameters (after rotation).}
\end{subfigure}
\begin{subfigure}[b]{1\textwidth}
\centering
\includegraphics[width=1\linewidth]{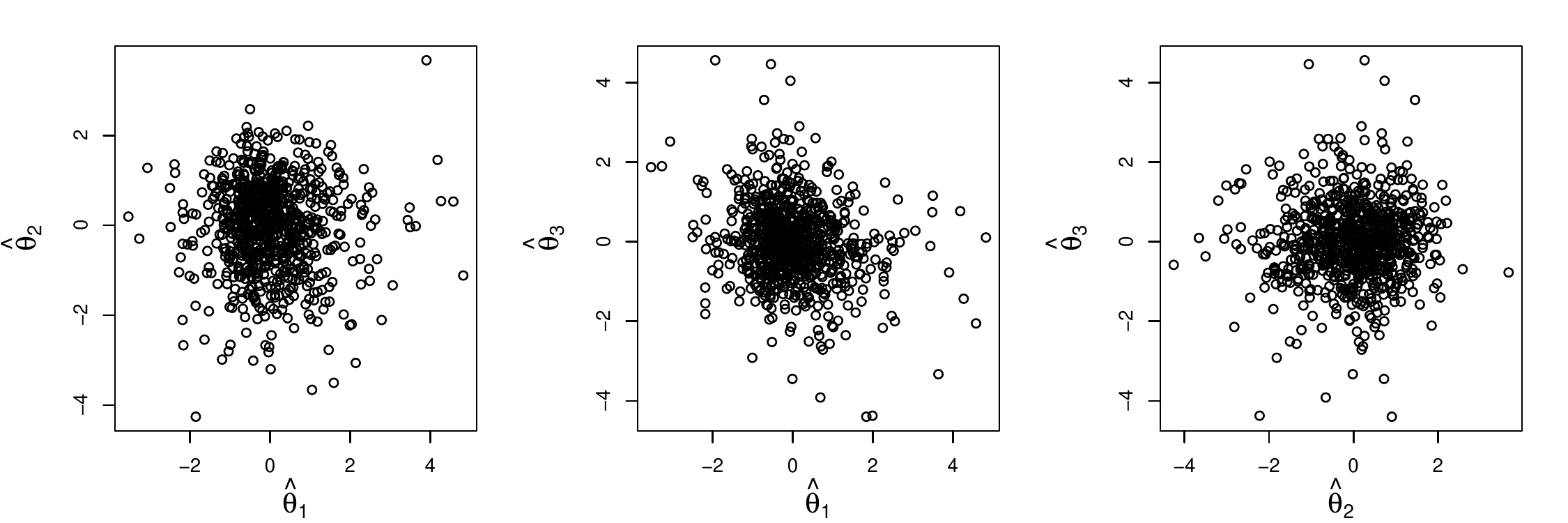}
 \caption{Scatter plots of the estimated person parameters (after rotation).}
\end{subfigure}
\caption{Fitting a three-factor model to the EPQ-R data: Histograms of the estimated person parameters (after Geomin rotation) of each dimension, and scatter plots of the estimated person parameters (after Geomin rotation) for each pair of dimensions. }\label{fig:thetadist}
\end{figure}

We further investigate the estimated person parameters.
In Figure~\ref{fig:thetadist}, we show the histograms of the estimated person parameters of each dimension, as well as the scatter plots of the estimated person parameters for each pair of dimensions. According to the histograms, the estimated person parameters on each dimension seem to be unimodal and almost symmetric about the origin. In addition, no obvious person clusters are found according to the scatter plots.
Table~\ref{tab:theta2_cor} further shows the correlations between the three estimated factors (after rotation). These correlations are relatively low, suggesting that Eysenck's three personality factors in Eysenck's model tend to be independent of each other.

Finally, a complete table of the estimated loading parameters is provided in the supplementary material.



\begin{table}
  \centering
  \footnotesize
  \begin{tabular}{c|rl}
    \hline
    Factor &Items  & Content \\
    \hline
     &35(E$+$)&Are you rather lively?\\
     &53(E$-$)&Do you tend to keep in the background on social occasions?\\
F1   &52(E$+$)&Do other people think of you as being very lively?\\
     &44(E$+$)&Do you like mixing with people?\\
     &42(E$+$)&Can you easily get some life into a rather dull party?\\
     \hline
     &1(P$+$)&Would you take drugs which may have strange or dangerous effects?\\
     &21(P$-$)&Are good manners very important?\\
F2   &7(P$+$)&Do you think marriage is old-fashioned and should be done away with?\\
     &22(P$-$)&Do good manners and cleanliness matter much to you?\\
     &12(P$+$)&Would you like other people to be afraid of you?\\
     \hline
     &64(N$+$)&Are you a worrier?\\
     &73(N$+$)&Do you worry too long after an embarrassing experience?\\
F3   &56(N$+$)&Does your mood often go up and down?\\
     &58(N$+$)&Do you often worry about things you should not have done or said?\\
     &61(N$+$)&Do you often feel `fed-up' ?\\
     \hline
\end{tabular}
\caption{Fitting a three-factor model to the EPQ-R data: The top five items with highest absolute loadings on each factor, under the Geomin rotation.}\label{tab:3factor}
\end{table}

\begin{table}
\centering
\begin{tabular}{c|ccc}
\hline
   & $\hat\theta_1$ & $\hat\theta_2$ & $\hat\theta_3$\\
 \hline
$T_1$ &  0.89 & 0.06 & -0.20 \\
$T_2$ &  0.11 & 0.88 & -0.02 \\
$T_3$ & -0.14 & 0.10 &  0.95 \\
\hline
\end{tabular}
\caption{Fitting a three-factor model to the EPQ-R data:  The correlations between the estimated person parameters and the corresponding total scores on the three scales. The rows of the table  ($T_1$, $T_2$, $T_3$) correspond to the total scores on the three scales and the columns ($\hat \theta_1$, $\hat \theta_2$, $\hat \theta_3$,) correspond to the estimated person parameters (after Geomin rotation). }\label{tab:score_cor}
\end{table}

\begin{table}
\centering
\begin{tabular}{c|ccc}
\hline
   & $\hat\theta_1$ & $\hat\theta_2$ & $\hat\theta_3$ \\
 \hline
$\hat\theta_1$ &  1.00 & -0.02 & -0.21 \\
$\hat\theta_2$ & -0.02 &  1.00 &  0.03 \\
$\hat\theta_3$ & -0.21 &  0.03 &  1.00 \\
\hline
\end{tabular}
\caption{Fitting a three-factor model to the EPQ-R data: The correlations between the estimated person parameters (after Geomin rotation).}\label{tab:theta2_cor}
\end{table}

\section{Discussion}\label{Sec:disc}

In this paper, we develop a statistical theory of joint maximum likelihood estimation
under an exploratory item factor analysis framework. In particular, a constrained joint maximum likelihood estimator is proposed that differs from the traditional joint maximum likelihood estimator by adding constraints on the Euclidian norms of both the item-wise and person-wise parameters. It is shown that this estimator consistently recovers the person and item specific response probabilities and also consistently estimates the loading matrix up to a rotation, under an asymptotic regime when both the numbers of participants and items grow to infinity.


An efficient alternating minimization algorithm is proposed for the computation that is scalable to large datasets with tens of thousands of people, thousands of items, and more than ten latent traits.
This algorithm iterates between two steps: updating person parameters given item parameters and updating item parameters given person parameters. In each step, the parameters can be updated in parallel for different people/items.
A novel projected gradient descent update is used in each step to handle the constraints.
Both our theory and computational methods are extended to analyzing data with missing responses.



The proposed method may be extended along several directions. First, the proposed theory and methods will be extended to
IFA models for polytomous response data which are commonly encountered in practice.
Specifically, we believe that similar theoretical results can be established for multidimensional graded models \citep[e.g.,][]{cai2010high}. More precisely, in a multidimensional graded model with $K$ factors, the latent structure is still reflected by a $J\times K$ loading matrix. This loading matrix should still be consistently recovered by a CJMLE, under the same asymptotic regime.

Second, even after applying rotational methods, the obtained factor loading matrix may not be simple (i.e., sparse) enough for a good interpretation. To better pursue a simple loading structure, it may be helpful to further add $L_1$ regularization of factor loading parameters \citep{sun2016latent} into the current optimization program for CJMLE, under which the estimated factor loading matrix is  automatically sparse and thus no post-hoc rotation is needed.
The statistical consistency of this $L_1$ regularized CJMLE may be further established, for which the issue of rotational indeterminacy may disappear.

Third, the missing responses are assumed to be missing completely at random in our theoretical analysis of missing data. As mentioned earlier, we believe that similar asymptotic properties still hold when relaxing this assumption to missing at random. This is left for future investigation.

Fourth, the current theoretical framework requires the number of latent factors to be known. When it is unknown, we suggest a cross-validation approach for choosing the latent dimension, which turns out to perform well according to our simulation studies and real data analysis. The statistical properties of this approach remain to be investigated. Alternatively, information criteria may be developed for determining the latent dimension.

\yc{In summary, this paper is a call to change the stereotype of joint maximum likelihood estimation as a statistically inconsistent method and a call to draw researchers' attention to the development of theory and methods for JML-based estimation.
JML-based estimation is generally applicable to almost all latent variable models,  easy to program, and computationally efficient. 
We believe that with a better theoretical understanding, JML-based estimation may become a new paradigm for the statistical analysis of latent variable models, especially for the analysis of complex and large-scale data.}


\bibliographystyle{apa}

\end{document}